\documentclass[aps,twocolumn,float,superscriptaddress,prd,nofootinbib]{revtex4-1}
\usepackage{graphicx, epsfig, bm, amsmath}
\usepackage{color}
\usepackage{hyperref}
\usepackage{ wasysym }
\usepackage{array}

\begin{document}

\newcommand{\lcdm}{$\Lambda$CDM}
\newcommand{\gpr}{G^{\prime}}
\newcommand{\fnl}{f_{\rm NL}}
\newcommand{\curv}{{\cal R}}
\newcommand{\R}{\mathcal{R}}
\newcommand{\dotR}{\dot{\mathcal{R}}}
\newcommand{\ddotR}{\ddot{\mathcal{R}}}
\newcommand{\ep}{\epsilon_H}
\newcommand{\dotep}{\dot{\epsilon}_H}
\newcommand{\et}{\eta_H}
\newcommand{\dotet}{\dot{\eta}_H}
\newcommand{\cs}{c_s}
\newcommand{\esq}{\left(}  % esq = esquerda = left
\newcommand{\dir}{\right)} % dir = direita  = right
\newcommand{\ord}{\mathcal{O}}
\newcommand{\hR}{\hat{\mathcal{R}}}
\newcommand{\dothR}{\dot{\hat{\mathcal{R}}}}
\newcommand{\mpl}{M_{\text{pl}}}

\definecolor{darkgreen}{cmyk}{0.85,0.2,1.00,0.2}
\newcommand{\vin}[1]{\textcolor{darkgreen}{[{\bf VM}: #1]}}
\newcommand{\wh}[1]{\textcolor{blue}{[{\bf WH}: #1]}}

\title{Inflationary Steps in the Planck Data}

\author{  Vin\'icius Miranda }
\affiliation{Department of Astronomy \& Astrophysics, University of Chicago, Chicago IL 60637}
\affiliation{The Capes Foundation, Ministry of Education of Brazil, Bras\'ilia DF 70359-970, Brazil}
\author{ Wayne Hu}
\affiliation{Kavli Institute for Cosmological Physics,  Enrico Fermi Institute, University of Chicago, Chicago, IL 60637}
\affiliation{Department of Astronomy \& Astrophysics, University of Chicago, Chicago IL 60637}

\begin{abstract}
We extend and improve the modeling and analysis of large-amplitude, sharp inflationary steps for second order corrections required by the precision of the
Planck CMB power spectrum and for  arbitrary Dirac-Born-Infeld  sound speed.    With two parameters, the amplitude and frequency of the resulting oscillations,
step models improve the fit by $\Delta \chi^2 = -11.4$.  Evidence for oscillations damping 
before the Planck beam scale is weak: damping only improves the fit to
$\Delta \chi^2 = -14.0$ for one extra parameter, if step and cosmological parameters
are jointly fit, in contrast to analyses
which fix the latter.  Likewise, further including 
the sound speed as a parameter only marginally improves the
fit to $\Delta \chi^2 = -15.2$ but has interesting implications for the lowest multipole 
temperature and polarization anisotropy.   Since chance features in the noise can mimic
these oscillatory features, we discuss tests from polarization power spectra, lensing 
reconstruction and squeezed and equilateral bispectra that should soon verify or 
falsify their primordial origin.
\end{abstract}

\maketitle

\section{Introduction}

Intriguingly, the cosmic microwave background (CMB) seems to favor rapid oscillations
in the curvature power spectrum over the smooth power law spectrum given by slow-roll
inflation at a level of $\Delta\chi^2 \sim 10-20$.   Such oscillations, first seen in the WMAP data \cite{Flauger:2009ab,Adshead:2011jq,Peiris:2013opa,Meerburg:2013cla}, persist in the recent Planck data \cite{Ade:2013rta,Easther:2013kla,Benetti:2013cja,Meerburg:2013dla}.
While the  significance of this improvement is debatable given the ability of
statistical fluctuations from instrument noise or cosmic variance to mimic the signal,
its implications for inflationary physics are sufficiently dramatic to merit careful 
consideration.

Rapidly oscillating power spectra can be generated during inflation if
the inflaton rolls over features in much less than an efold, for example 
oscillations in the potential \cite{Chen:2006xjb,Silverstein:2003hf},
a step in the potential \cite{Adams:2001vc} or warp in the Dirac-Born-Infeld (DBI) 
model \cite{Bean:2008na}.   
In this paper we consider the less well-explored step feature cases.

On the model side, we extend previous analyses  \cite{Adshead:2011jq,Miranda:2012rm}
by analytically treating large amplitude sharp steps in both the potential and warp at 
arbitrary sound speeds including new second order corrections that are 
required by the enhanced precision of the Planck data.  Having an analytic model for the inflationary power spectrum greatly enhances the efficiency of  the analysis while  varying the sound speed
provides interesting phenomenology for the lowest multipoles. 

 On the analysis side, 
we jointly fit for step and cosmological parameters unlike the Planck collaboration
analysis \cite{Ade:2013rta}.   Because the presence of step oscillations also changes
the broadband average power in the spectrum, joint variation is crucial for interpreting
constraints on step parameters.
Although more recent analyses have also jointly
varied parameters \cite{Easther:2013kla,Meerburg:2013dla}, they did so in a
different context where the oscillations persist out to arbitrarily high multipoles.
We show that joint variation is particularly important for finite width steps and  
misleading constraints arise when cosmological parameters are fixed. 

The outline of the paper is as follows.  In \S\ref{sec:step} we describe the improvements and extensions to the modeling
of the curvature power spectrum from steps in the potential and warp.  These  are derived in
Appendix \ref{app:analytic} and shown to be sufficiently accurate for Planck data in Appendix \ref{sec:accuracy}.   The best fit step models at low and high sound speed, found
from jointly maximizing the likelihood over step and cosmological parameters
 in Appendix~\ref{sec:minimize},  are
presented in \S\ref{sec:planck}.  In \S \ref{sec:future} we provide falsifiable predictions
of these models.   We discuss these results in \S\ref{sec:discussion}.

\section{Step Power Spectra}
 \label{sec:step}
 
In this section, we summarize the description of the curvature power spectrum for sharp potential
and warp steps in DBI inflation derived in Appendix \ref{app:analytic}. This analytic
treatment generalizes previous
ones \cite{Adshead:2011jq,Miranda:2012rm} to large amplitude, arbitrary sound speed
models and employs second order corrections to ensure sufficient accuracy for 
comparison to the Planck data in the following sections.

We consider models with step features in the DBI Lagrangian
\begin{equation}
{\cal L}= \left[
1-\sqrt{1 - 2  X/T(\phi)} \right] T(\phi)- V(\phi),
\label{eqn:DBI}
\end{equation}
where the kinetic term  $2X = - \nabla^{\mu} \phi \nabla_{\mu} \phi$.  
We choose units where $M_{\rm pl}=(8\pi G)^{-1/2} = c = \hbar = 1$ throughout.
In braneworld theories that motivate the DBI Lagrangian, $\phi$ determines the position of the brane, $T(\phi)$ gives the warped brane tension, and $V(\phi)$ is the interaction potential.  
Note that for $X/T \ll 1$, the sound speed
\begin{equation}
c_s(\phi,X) =\sqrt{ 1 - 2 X/T(\phi)},
\end{equation}
goes to 1 and
 this Lagrangian becomes that of a scalar with a canonical
kinetic term.

We allow steps to 
appear in either the warp or the potential
\begin{align}\label{eqn:T(phi)}
	T(\phi) =& \frac{\phi^4}{\lambda_B} [1 + b_T F(\phi) ],  \nonumber\\
V(\phi) =&  V_0 \Big(1 - \frac{1}{6}\beta \phi^2\Big) [1 + b_V F(\phi) ].
\end{align}
 Here $\lambda_B$, $V_0$, $\beta$ parameterize
the smooth model and are determined by the tilt and amplitude of the power spectrum as
well as the end point for DBI inflation, 
whereas $b_T$, $b_V$ give the height of a $\tanh$ step
\begin{align}
	F(\phi)= \tanh\Big(\frac{\phi-\phi_s}{d}\Big)-1,
\end{align}
at field location $\phi_s$, with field width $d$.  Unlike previous treatments  \cite{Adshead:2011jq,Miranda:2012rm}
we allow
for the possibility of potential steps at arbitrary sound speed but for simplicity do not
consider simultaneous steps in both the warp and the potential.

We show in Appendix \ref{app:analytic} that steps in the warp or potential, over which the inflaton rolls in much less than an efold,  generate
 oscillations in the power spectrum of the following form
\begin{align} \label{eqn:ps_analytical_form}
	 \ln \Delta_{\mathcal{R}}^2 = & \ln  A_s \left(\frac{k}{k_0}\right)^{n_s-1}  +I_0(k) + 
	 \ln [1+ I_1^2(k)] ,
\end{align}
where we take 
%the normalization scale $k_0=0.05$ Mpc$^{-1}$.
the normalization scale $k_0=0.08$ Mpc$^{-1}$ which is closer to
the best constrained scale for the Planck data than the conventional choice of $0.05$ Mpc$^{-1}$.
The leading order contribution from the step is
\begin{align}
I_0(k)= \Big[C_{1} W(k s_s) + C_{2} W'(k s_s) %\nonumber \\ &
 + C_{3} Y(k s_s) \Big]  \mathcal{D}\Big(\frac{k s_s}{x_d}\Big), 
	\end{align}	
and the second order contribution is
\begin{align}
\sqrt{2}I_1(k) =& \frac{\pi}{2}(1-n_s) +\Big[ C_1 X (k s_s) + C_2 X' (k s_s)  \nonumber\\&
+ C_3 Z (k s_s)  \Big]  \mathcal{D}\Big(\frac{k s_s}{x_d}\Big) ,
\end{align}
where the windows  
\begin{align}
	\label{eqn:powerwindow}
W(x) &= {3 \sin(2 x) \over 2 x^3} - {3 \cos (2 x) \over x^2} - {3 \sin(2 x)\over 2 x} , \nonumber\\
X(x) & = {3 \over x^3} (\sin x - x \cos x)^2 , \\
	Y(x) &= \frac{6x\cos(2x) + (4x^2-3) \sin(2x)}{x^3}, \nonumber\\
	Z(x) &= -\frac{3+2x^2-(3-4x^2)\cos(2x)-6x\sin(2x)}{x^3}, \nonumber
\end{align}
and  $'={d/d\ln x}$.
The sound horizon when the inflaton
crosses the step $s_s$ controls the frequency of the oscillations, whereas the finite width $x_d \propto d^{-1}$ 
determines their damping via 
\begin{equation}
\mathcal{D}(y) = \frac{y}{\sinh(y)} .
\end{equation}   
We give the correspondence between these phenomenological parameters and the fundamental ones in Appendix \ref{sec:accuracy}.   There we also test the accuracy of the
analytic model in Eq.~(\ref{eqn:ps_analytical_form}) against exact calculations.   We show
that the precision of the Planck data set necessitates the inclusion of the second order
$I_1$ correction whose analytic form is entirely new to this work (cf.~\cite{Miranda:2012rm},\cite{Ade:2013rta}).  Note that the second order term is determined by exactly the
same parameters as the leading order term as a consequence of the generalized slow-roll  construction
\cite{Dvorkin:2010dn}.

The constants $C_i$ can be related to fractional changes in $c_s$ and the slow roll parameter $\epsilon_H=-d\ln H/d\ln a$ induced 
 by the step
\begin{align}
c_j \equiv & \frac{c_{sj}}{c_{sa}}, \quad e_j \equiv  \frac{\epsilon_{Hj}}{\epsilon_{Ha}} ,
\end{align}
where  $``a"$ denotes their values on the attractor after the step,
 $``b"$ for the same before the step, $``i"$ for immediately after the step off of the attractor.
 More specifically,
\begin{align} \label{eqn:CW}
    C_{1} =& -\ln c_b e_b , \nonumber\\
    C_{2} =& -\frac{2}{3} \frac{c_i-c_b}{c_i + c_b}  + \frac{2}{3} \frac{e_i-e_b}{e_i+e_b},\nonumber\\
    C_{3} =& 2\frac{(1-c_b) +(c_i-1)/4}{c_i+c_b}.
\end{align}
For  warp steps,  the attractor solutions before and after the step and energy
conservation at the step gives $ (b_V=0)$
\begin{align}
c_b &=  e_b =\sqrt{\frac{1-2b_T}{1-2b_T c_{sa}^2}},\nonumber\\
c_i &= \frac{c_b }{1- 2b_T(1-c_{sa} c_b)} ,\nonumber\\
e_i &=  \frac{1- c_{sa}^2 c_i^2}{c_i(1-c_{sa}^2)}.
\end{align}
 This generalizes the results of Ref.~\cite{Miranda:2012rm} to large
amplitude steps as $c_{sa} \rightarrow 1$.  Note that in this limit, arbitrarily large
fractional steps in the warp $b_T \rightarrow -\infty$ still only cause infinitesimal
changes in the slow roll parameters $c_j=e_j=1$ or $C_i \rightarrow 0$.  Consequently, there are sound speeds near unity for which a step in the warp cannot
explain finite amplitude oscillations in the data.

\begin{figure}[t]  
\psfig{file=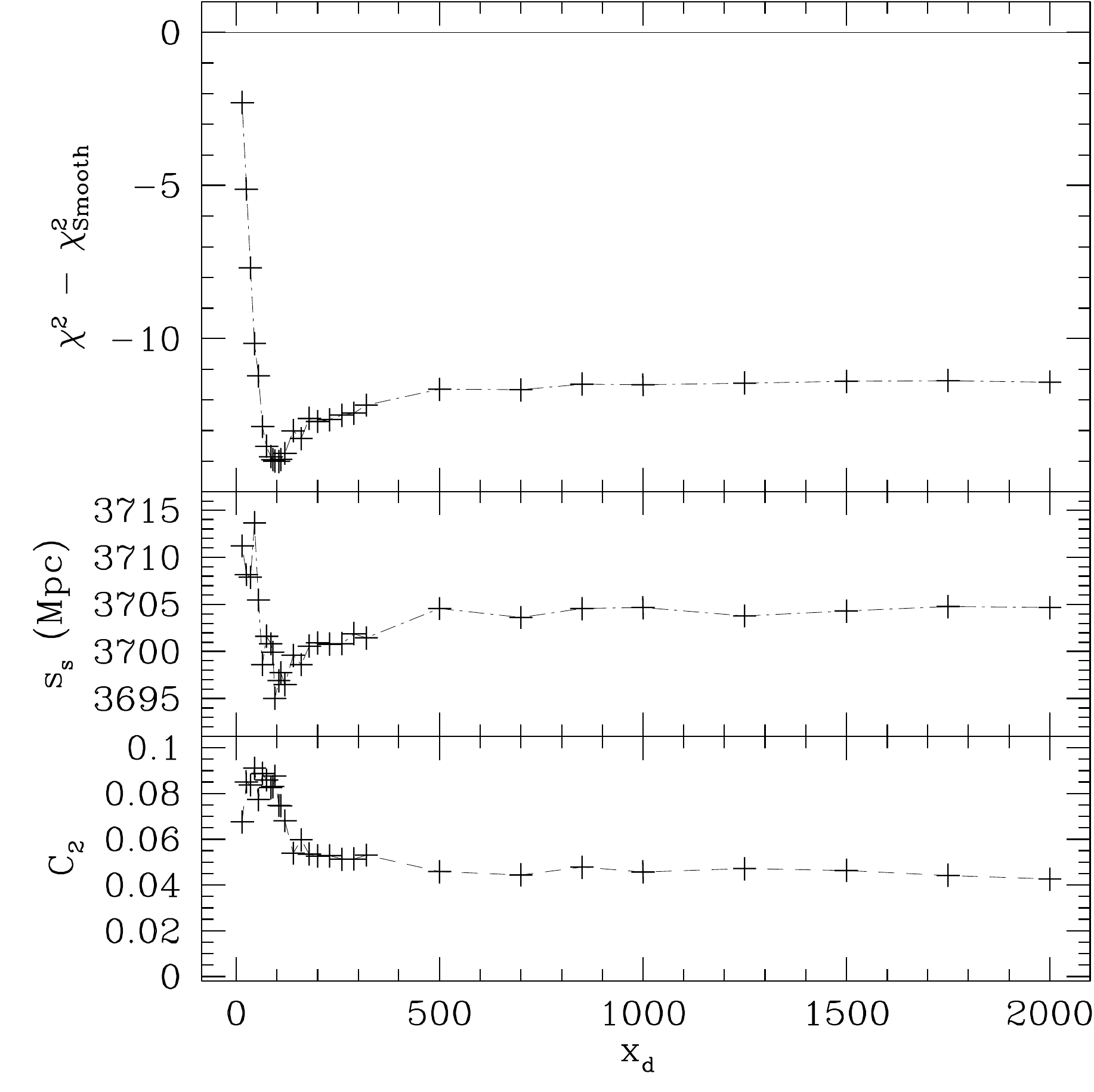, width=3.25in}
\caption{Minimum $\chi^2$ relative to the best fit smooth (no step) model 
as a function of the oscillation damping scale $x_d$ for the
$c_s=1$ potential step model (top panel).   The minimization is performed jointly over
cosmological  and step parameters with  step position $s_s$ and amplitude of  oscillations $C_2$ shown in the middle and bottom panels respectively.} \label{plot:minplanck_GSR0_GSR1}	
\end{figure}

For potential steps $(b_T=0)$
\begin{align}
c_b=&e_b=1, \nonumber\\
c_i = &1-\frac{ 3 b_V (1- c_{sa}^2)}{3 b_V (1- c_{sa}^2) - \epsilon_{Ha}},\nonumber\\
e_i =& 1 -\frac{ 3 b_V [-3b_V(1- c_{sa}^2) + (1+c_{sa}^2) \epsilon_{Ha}]}{\epsilon_{Ha} [-3 b_V(1- c_{sa}^2)+ \epsilon_{Ha} ]}.
\end{align}
This generalizes the results of Ref.~\cite{Adshead:2011jq} for potential steps to arbitrary
sound speeds.  Note that for potential steps $C_1=0$.  We test the accuracy of these approximations 
in Appendix \ref{sec:accuracy}.

In summary, 
a model is  parameterized by five numbers $\{ C_1,C_2,C_3, s_s,x_d \}$.  
To leading order, the $C_1$ term represents a step in the power spectrum at $k  s_s \sim 1$,
the $C_2$ term represents a constant amplitude oscillation out to the damping scale $k s_s=x_d$, and
the $C_3$ term represents a change to the shape of the first few oscillations.

In a given model,
not all of these parameters are independent as they are determined by the background
parameters.   Specifically, the three $C_i$ parameters are controlled by
the amplitude of the step and mainly the sound speed after the step.  
For a potential step $C_1=0$ and
$C_3\rightarrow 0$ for $c_s \rightarrow 1$.   For a warp step, all three $C_i$ are comparable but $C_i \rightarrow 0$ for $c_s \rightarrow 1$, even for arbitrarily large steps.

\section{Planck Data Analysis}
\label{sec:planck}

In this section we analyze the Planck data for the presence of
sharp inflationary steps which create high frequency oscillations in the power spectrum.   We begin in \S \ref{sec:potsteps} with potential steps in
canonical sound speed models.   In \S \ref{sec:lowsound} we extend the analysis
to arbitrary sound speed models where both warp and potential steps can produce the
oscillatory phenomenology favored by the Planck data.    In Appendix \ref{sec:minimize}
we discuss details of the analysis that enhance the efficiency of the model search.

\begin{figure}[t]
\psfig{file=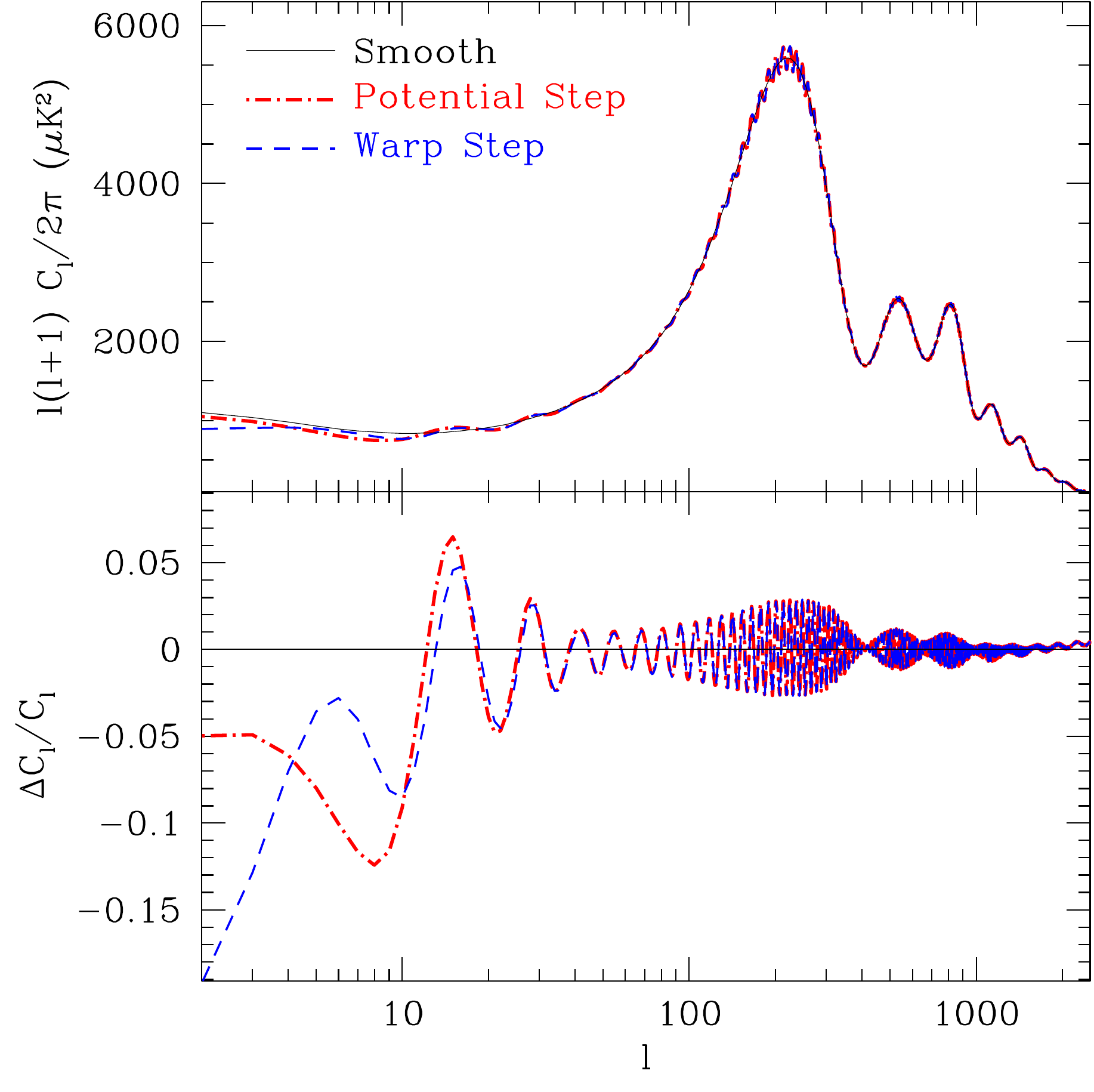, width=3.25in}
\caption{Best fit  models for a
potential step with $c_{sa} = 1.0 $ (red) and a warp step at $c_{sa} = 0.7$ (blue).  Both
models have the same few percent oscillations around the best fit smooth model at the
first and second peaks (lower panel).  The warp step is a marginally better fit of
$\Delta\chi^2 = -15.2$, versus the potential step with $\Delta\chi^2 = -14.0$, due to 
suppression of the lowest multipoles but introduces the sound speed as an additional parameter. }
\label{fig:cs08_ps}
\end{figure}	

\subsection{Canonical Sound Speed Models}
\label{sec:potsteps}

We begin with the simplest case of $c_s=1$ models.   Here, warp steps have no effect and potential steps give $C_1=C_3=0$.  Step models are thus  described by three parameters $\{ s_s, C_2, x_d \}$, the oscillation
frequency, amplitude and damping scale respectively.  
 The underlying smooth cosmology
is taken to be the  flat $\Lambda$CDM model as defined by $\{A_s, n_s,\theta_A, \Omega_c h^2, \Omega_b h^2, \tau\}$, where $\theta_A$ is the angular acoustic scale at recombination, $\Omega_c h^2$ parameterizes the cold dark matter density, $\Omega_b h^2$ the baryon density, $\tau$ the Thomson optical depth to reionization.
 We calculate CMB power spectra
using a modified version of CAMB \cite{Lewis:1999bs,Howlett:2012mh}. 
In addition, the Planck data are modeled by foreground parameters which we hold fixed
throughout to the best fit smooth model (see Appendix \ref{sec:minimize} and
Tab.~\ref{table:smooth_parameters}, \ref{table:foregrounds}).   

We are interested in the question of whether the step parameters significantly improve the 
fit to the Planck data rather than marginalized constraints on the parameters themselves.
Since the Monte Carlo Markov chain technique is highly inefficient for these purposes, we instead directly maximize the likelihood or minimize the effective $\chi^2 =-2\ln {\cal L}$ 
in the  step and cosmological parameter space jointly.   For these 
oscillatory spectra, the likelihood is a rapidly varying function of frequency $s_s$ with
many local minima.   Fortunately previous works have  shown that $s_s \approx 3700\,\text{Mpc}$ is the frequency range that contains the global minimum \cite{Ade:2013rta,Meerburg:2013dla}.
We therefore search only around this global minimum region.  Even so, for efficiency in
the minimization it is important
to choose combinations of the parameters that are close to the principal components
of the curvature or covariance matrix.  We discuss such choices in Appendix \ref{sec:minimize}.

Unlike the Planck collaboration analysis \cite{Ade:2013rta}, we simultaneously vary the cosmological and step parameters
in the minimization.  This step is crucial as discussed in Appendix \ref{sec:accuracy}
since the presence of rapid oscillations also changes both the amplitude and shape of the 
broadband power in multipole space.   If the cosmological parameters are not
readjusted, the Planck data would falsely suggest that the oscillations cannot continue 
into the $\ell \sim 10^3$ regime where the data is most constraining.
  For this reason, the minimum found in Ref.~\cite{Ade:2013rta} is not the global minimum nor is there strong evidence for damping of the 
oscillations at high multipole.    The minimum $\chi^2$ as a function of the
damping scale $x_d$ is shown in Fig.~\ref{plot:minplanck_GSR0_GSR1} and is nearly flat
for $x_d > 10^2$.

We find that the global minimum is given by
\begin{align} \label{eqn:best_fit_canonical_model}
	 C_2 &= 0.075, \nonumber\\
	s_s  &= 3696.9\, {\rm Mpc}, \nonumber\\
	x_d &= 105.0, \nonumber\\
	\Delta \chi^2 &= -14.0,
\end{align} 
where the $\chi^2$ improvement is measured against the best fit smooth model.  
In contrast, the best fit model of Ref.~\cite{Ade:2013rta} had $x_d = 87$ and a 
similar amplitude and frequency with a $\Delta\chi^2=-11.7$.\footnote{The original version of the
Planck collaboration analysis \cite{Ade:2013rta} [arXiv:1303.5082v1] erroneously conflated $C_2$ with $A_c$ (see Eq.~\ref{eqn:Ac_def}) and correcting this definition also brings the amplitude
into agreement with that found for WMAP \cite{Adshead:2011jq}.}  As shown in 
Tab.~\ref{table:representative_models}, this difference is not due to the inclusion of 
second order corrections from $I_1$ in Eq.~(\ref{eqn:ps_analytical_form}) though their
omission would bias cosmological parameters such as $A_s$ and $n_s$.
 The Planck data
thus favor oscillations at the few percent level in $C_\ell$, with a peak-to-peak spacing of
$\Delta \ell \sim 12$.    Note that at $k s_s = x_d$,  the damping suppression ${\cal D}=0.85$ and
for this model corresponds to $ \ell_d \approx 400$.   Thus the oscillations 
persists at least out to the
second acoustic peak (see Fig.~\ref{fig:cs08_ps}).   

	The complete list of cosmological parameters for the best fit step model is listed in Tab.\ref{table:representative_models}. The oscillations add broadband power and generally require
	a lower normalization.   Because in this
	model they damp near the well-constrained third peak, this also requires a higher tilt
	to keep the total power fixed at this best constrained region.
	As a result, the model has slightly smaller broadband power at low multipoles, reaching  $\sim -5\%$ at the quadrupole.

Furthermore, the Planck data are also compatible with oscillations that persist
out to the highest multipole measured in the data $\ell=2500$.  
Taking $x_d \approx 2000$, which is indistinguishable from infinity for Planck, 
\begin{align}
	 C_2 &= 0.043, \nonumber\\
	s_s  &= 3704.7\, {\rm Mpc}, \nonumber\\
%	x_d &= 2000, \nonumber\\
	\Delta \chi^2 &= -11.4.
\end{align}
This fit only differs in $\chi^2$ by 2.6 from that of the  global minimum and is comparable
to the best fit found in Ref.~\cite{Ade:2013rta}.  Note that even with no damping
scale in curvature fluctuations, oscillations in $C_\ell$ decline with $\ell$ due to 
projection and lensing effects (see \cite{Adshead:2011jq} and  Eq.~\ref{eqn:Ac_def}).
Had we fixed cosmological parameters
to the best fit smooth model then this $x_d$ would be falsely 
penalized by $\Delta \chi^2 = 9$.   
The cosmological parameters for this model are
also given in Tab.~\ref{table:representative_models}.  Notably with no damping scale,
the tilt no longer requires significant adjustment.  The change in $s_s$ mainly reflects
slightly different cosmological parameters that produce correspondingly different distances to
recombination rather than a change in the angular scale.

In summary, the Planck data favor percent level oscillations in $C_\ell$ produced by potential step features by $\Delta \chi^2 = -11.4$ with
two parameters that control the oscillation frequency and amplitude.    Minimizing the 
$\chi^2$ for damping of the oscillations confines the oscillations to roughly the first and
second peaks and marginally improves the fit with an additional
$\Delta \chi^2 = -2.6$ for a total of $-14.0$ with one additional parameter for a total of three.

\begin{figure}[t]
\psfig{file=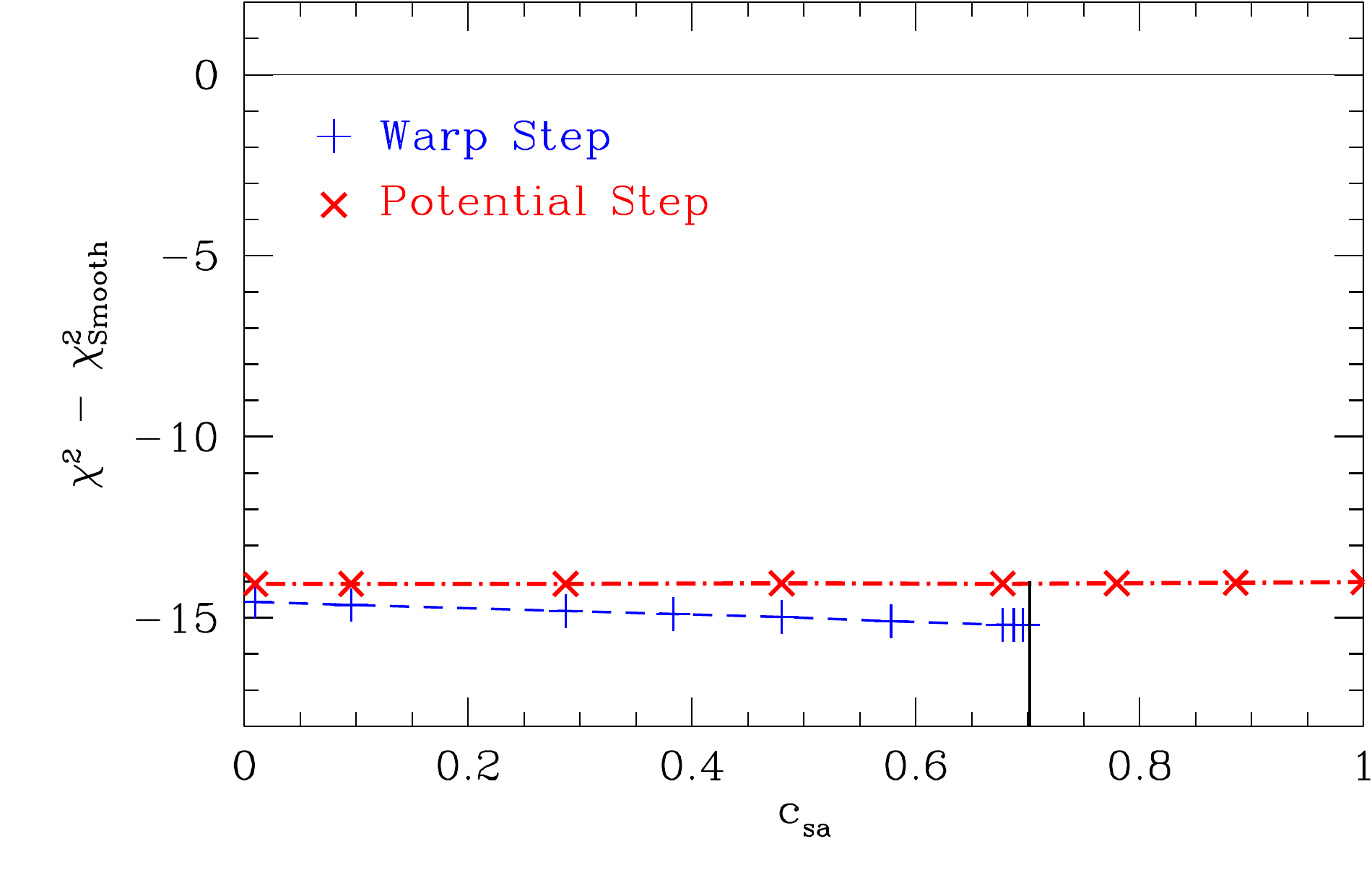, width=3.25in}
\caption{ $\chi^2$ improvement as a function of the sound speed after the step $c_{sa}$ for
the warp  (blue $+$) and potential (red $\times$) steps.   Other parameters are fixed
to their minimum $\chi^2$ values in the $c_s=1$ potential step model including the oscillation amplitude $C_2$. Note that
warp steps with $c_{sa} \gtrsim 0.7$ cannot generate the required $C_2$ (see Eq.~\ref{eqn::csmaxwarp}) as marked by the vertical line.}
\label{fig:C1C3DBI}
\end{figure}

%%%%%%%%%%%%%%%%%%%%%%%%%%%%%%%%%%%%%%%%%%%%%%%%%%%%%%%%%%%%%%%%%%%%%%%%%%%%%%%%%%%%%%%%%%%%%%%%

\subsection{Low Sound Speed Models}
\label{sec:lowsound}

Low sound speed DBI models allow for two different classes of steps with two different
phenomenologies that impact low multipoles in the Planck data.  Both steps in the
potential $V(\phi)$ and warp $T(\phi)$ produce the same high multipole oscillations
driven by the amplitude parameter $C_2$.    Given a $C_2$ that minimizes the Planck
$\chi^2$ at high multipole, the remaining freedom is in choosing a sound speed after
the step $c_{sa}$.   
  In both the potential and warp scenarios, this uniquely fixes
the two remaining step parameters $C_1$ and $C_3$.  Recall that $C_1$ controls
the step in the power spectrum around the first oscillation and $C_3$ controls the shape 
of the first few oscillations. 
Since the fit is driven by the $C_2$ oscillations with
only small impact from $C_1$ and $C_3$, we fix all the other parameters to the global
minimum of Eq.~(\ref{eqn:best_fit_canonical_model}) when examining the impact of $c_{sa}$.

For potential steps, $C_1=0$ and $-3/8 < C_3/C_2 < 0$. Even for the maximal
case of $-3/8$ and $c_{sa}\rightarrow 0$, there is very little impact on 
the CMB power spectrum.   Consequently as shown in Fig.~\ref{fig:C1C3DBI} the $\chi^2$
surface is essentially flat across $c_{sa}$.  

For warp steps both
$C_1/C_2$ and $C_3/C_2$ can be greater than unity and the sound speed has a larger 
fractional effect on $C_\ell$.  However, their impact is still limited to the first few
oscillations and, given the preference for a horizon scale $s_s$, severely cosmic 
variance limited.  Raising $c_{sa}$ mainly enhances the step in the power spectrum
relative to the oscillations thus lowering the first few multipoles.  Both $C_1$ and
$C_3$ are important in establishing the shape due to a cancellation in their effects
at the first oscillation.

Since warp steps do not produce oscillatory features as $c_s \rightarrow 1$, there
is a maximum $c_{sa} \sim 0.7$ for which they can explain the oscillations (see Fig.~\ref{fig:C1C3DBI}).
The best fit has the maximal possible sound speed
\begin{align}
	c_{sa} &= 0.70, \nonumber\\
	\Delta \chi^2 &= -15.2 \quad (\text{warp}),
\end{align}
which implies $C_1 = -0.70$,  $C_3 = -0.37$ given the fixed parameters in
Eq.~(\ref{eqn:best_fit_canonical_model}).  
While the step in the  curvature power  spectrum
is approximately $50\%$ (see Fig.~\ref{fig:test_analytics}) in $C_\ell$,
this and the changes in the cosmological parameters 
translates into a $\sim 20\%$ suppression of power at the quadrupole relative to the smooth model (see
Fig.~\ref{fig:cs08_ps}).  Note that the drop between $2 \le  \ell \le 5$ is particularly sharp
for warp steps due to a local maximum in the curvature spectrum oscillations.  Nonetheless,
with cosmic variance these changes have only a small impact on the fit.   
As a consequence, while warp steps have interesting phenomenology that may ameliorate low multipole anomalies,
there is no statistically significant preference for $c_s<1$.

\begin{figure*}[t]  
\psfig{file=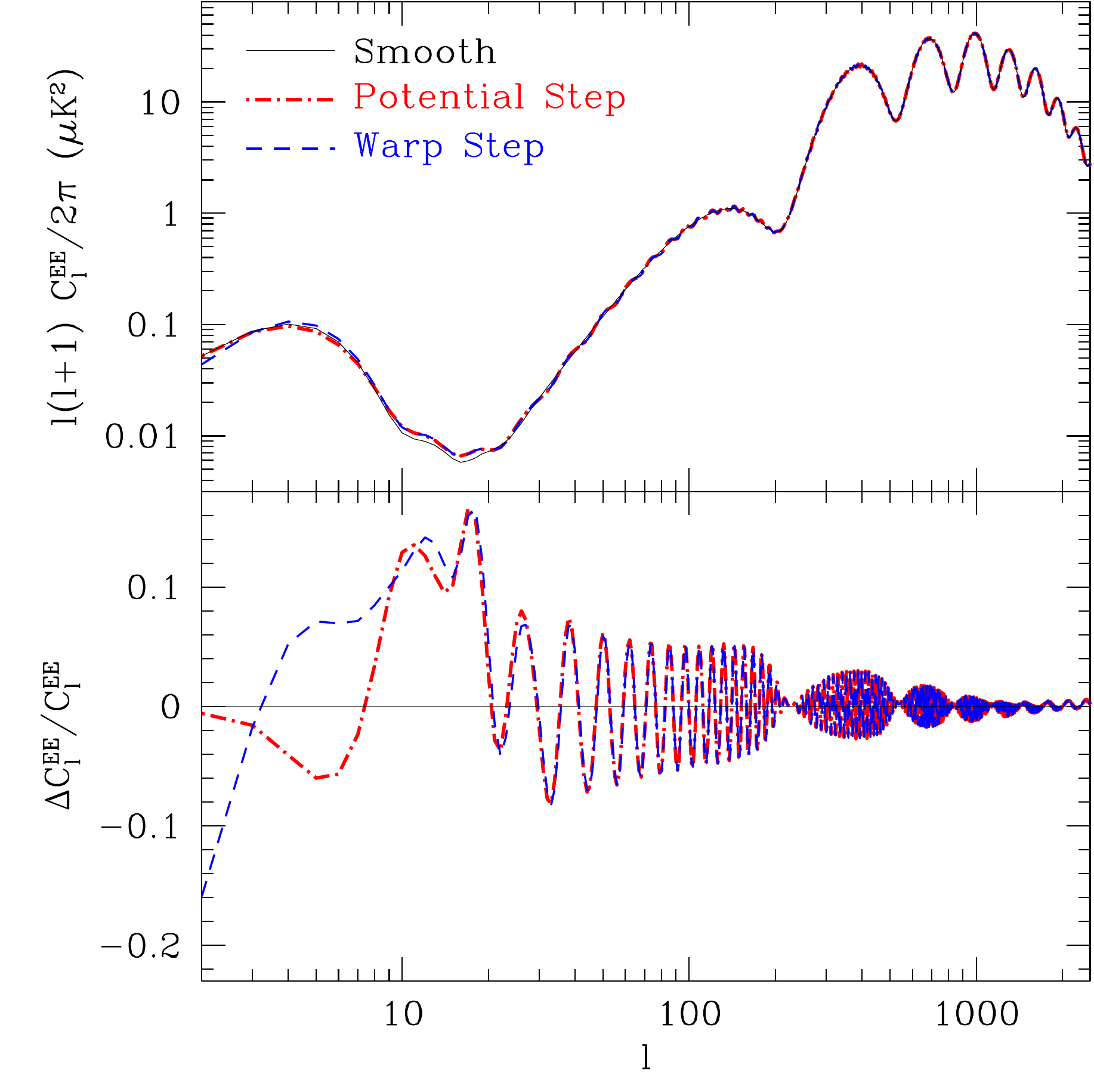, width=3.25in}\psfig{file=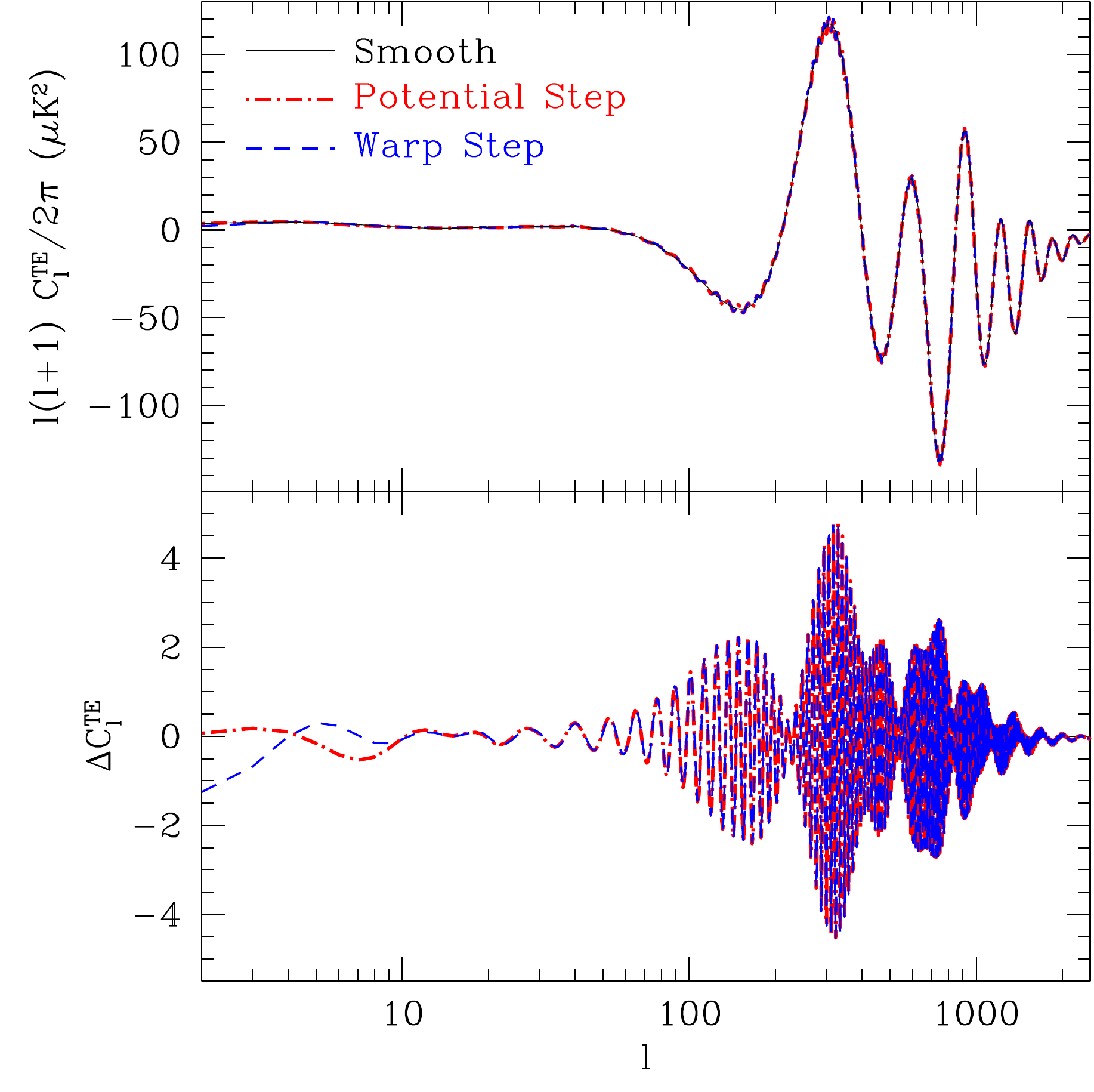, width=3.25in}
\caption{Polarization (left) and temperature polarization cross (right) power spectra
for the best fit models of Fig.~\ref{fig:cs08_ps}.  Step oscillations provide falsifiable predictions for the polarization which would not be mimicked by chance features in the noise.}
 \label{plot:pol_best_fit}	
\end{figure*}

\section{Future Tests}
\label{sec:future}

While an improvement of $\Delta \chi^2 \approx -11$ for two parameters and up to $-15$ in the
full step parameter space may sound significant, it has been shown that for more flexible oscillatory models, where
not only the amplitude and frequency but also the phase of the oscillation is fit, 
realizations of smooth models with noise often recover this level of improvement, albeit typically
with a smaller oscillation amplitude \cite{Meerburg:2013cla}.   Furthermore the improvement
in the WMAP likelihood \cite{Adshead:2011jq,Meerburg:2013cla} is comparable to that of the Planck likelihood
despite the higher precision of Planck whereas one would have expected the latter to 
increase for a true signal.   For these reasons, it is important to have more definitive
tests for the origin of these improvements.  In this section we discuss predictions of
the best fit models identified above that may be used to verify or falsify the hypothesis of their primordial
origin.

As emphasized by Ref.~\cite{Mortonson2009}, the most incisive consistency test for inflationary
features is the $E$-mode polarization power spectrum and cross spectrum.
  In
Fig.~\ref{plot:pol_best_fit} (left panel), we show the predicted $E$-mode power spectrum of 
the models in Fig.~\ref{fig:cs08_ps}.  Consistency with inflationary oscillations demands that 
 oscillations appear at the same frequency while modulated by the acoustic
 transfer to have nodes that are out of phase with the temperature.   Furthermore
 due to projection effects, the polarization oscillations are twice as prominent in polarization.
 In principle, the low $\ell$ polarization can also more than double the distinguishing
 power between the warp and potential fits, albeit limited in practice by galactic foregrounds
 and uncertainties in the reionization model.
 
 Finally, the temperature-polarization cross spectrum must also exhibit consistent oscillations
 as shown in Fig.~\ref{plot:pol_best_fit} (right panel).
 These predictions should be tested in the next release of the Planck data.  More 
 generally they can be tested in any CMB polarization data set that has sufficient amounts of sky to distinguish modes separated by $\Delta \ell \approx 12$ and oscillations in power of 3-10\%.

\begin{figure}[t]  
\psfig{file=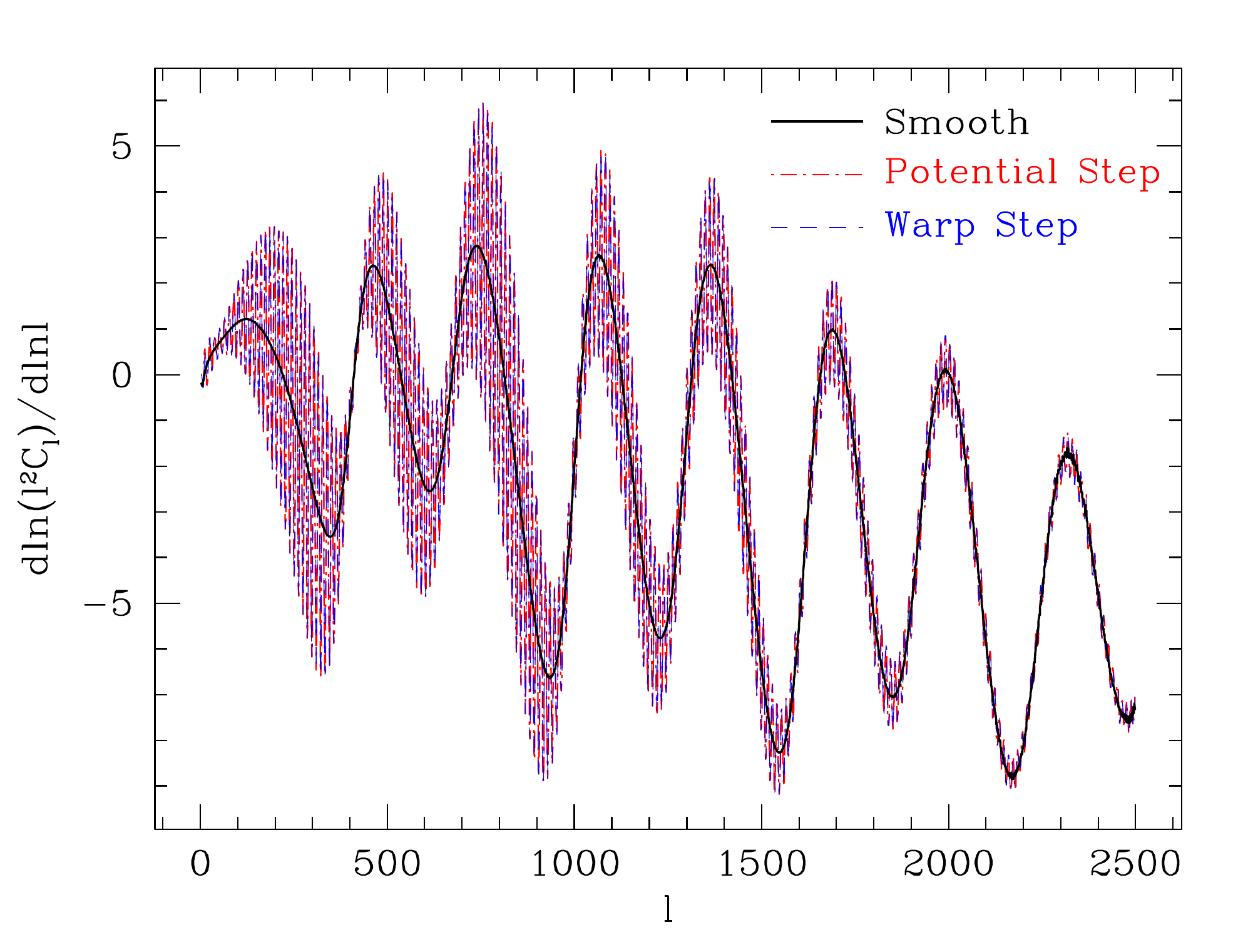, width=3.25in}
\caption{Temperature power spectrum derivatives for the best fit models of Fig.~\ref{fig:cs08_ps}.   Low amplitude, high frequency oscillations produce new signals for lensing reconstruction and  squeezed 
bispectra.}
 \label{plot:deriv}	
\end{figure}

The best fit oscillatory models also predict different CMB lensing effects.   
High frequency features act in a similar fashion as the acoustic peaks in providing
a signal for lensing.   The $\Delta \ell \approx 12$ fineness of the features compared
with the acoustic spacing of $\Delta \ell \approx 300$ offsets the smallness of the
amplitude.  In Fig.~\ref{plot:deriv}, we quantify this expectation by showing for the smooth and best fit step models
\begin{equation}
\frac{ d\ln \ell^2 C_\ell}{d\ln \ell},
\label{eqn:Clderivative}
\end{equation}
which controls lensing and squeezed bispectrum effects \cite{Lewis:2011fk}.  Note that what was a small effect for
the power spectrum is an order unity effect for certain lensing effects.

There are two related ways in which the oscillations impact lensing observables.
First, if lensing reconstruction were performed with the oscillatory models to 
construct the filters in the optimal quadratic estimator 
rather than the smooth models, the noise power in the reconstruction
should decrease if
the oscillatory features are real.   To see this note that in the flat-sky approximation the
sample-variance limited reconstruction noise power $N_L$ of the lensing potential is given by \cite{Hu:2001tn,Lewis:2011fk}
\begin{equation}
N_L^{-1} = \int \frac{d^2 \ell_1}{(2\pi)^2} \frac{({\bf L} \cdot {\boldsymbol \ell}_1 C_{\ell_1} + {\bf L} \cdot {\boldsymbol \ell_2} C_{\ell_2})^2}{2 C_{\ell_1} C_{\ell_2}},
\end{equation}
where ${\boldsymbol \ell}_1 + {\boldsymbol \ell}_2 = {\bf L}$.   For $L \ll \ell_1$, 
${\boldsymbol \ell}_1 \approx -{\boldsymbol \ell}_2$ and for $L \ll \Delta \ell$, $C_{\ell_2}$ can be Taylor expanded around $\ell_1$.    The numerator then scales as 
the derivative in Eq.~(\ref{eqn:Clderivative}) squared.     For the best 
fit models, this changes the noise for $L \lesssim 12$.

Relatedly, lensing by long-wavelength modes modulates the angular scale of
features in the power spectrum which itself is correlated with the CMB temperature
anisotropy through the ISW effect.   Thus the presence of fine scale oscillations
changes the squeezed reduced bispectrum of temperature fluctuations
\cite{Goldberg:1999xm}
\begin{align}
b_{L \ell_1\ell_2}& \approx \frac{ L(L+1)-\ell_1(\ell_1+1)+ \ell_2(\ell_2+1)}{2} 
C_L^{T\phi} C_{\ell_2} \nonumber\\
&\qquad + {5 {\rm perm.}},
\end{align}
where $C_L^{T\phi}$ is the correlation of the ISW temperature and lensing potential fields.
The permutation of $\ell_1 \leftrightarrow \ell_2$ again makes the result 
scale as the derivative of $C_\ell$ and is enhanced by the oscillation.  Both of these
lensing effects are in principle detectable, though for the best fit frequency with 
$\Delta \ell \sim 12$ the impact will be limited by cosmic variance.   A more detailed
study is required to determine their effects on the existing Planck data set.
%Both of these lensing effects can be tested in the existing Planck data set \cite{Ade:2013dsi,Ade:2013tyw}.

Features during inflation also  produce primordial non-Gaussianity in mainly the equilateral
configuration \cite{Chen:2006xjb,Adshead:2011bw,Adshead:2012xz,Achucarro:2012fd}.  For the 
best fit step models these should also be observable in Planck \cite{Adshead:2011jq,Adshead:2013zfa}.   Extracting these signals though will require using specific templates
that include these rapid oscillations
\cite{Fergusson:2010dm}.   Since the equilateral bispectrum amplitude scales 
as $x_d^2$, the lack of a strong bound on the damping scale implies that the
bispectrum signal could be very large at high multipole, though these models would be beyond the 
regime of validity of the effective field theory that underlies their calculation \cite{Baumann:2011su}
% does not seem to be about limitations of effective field theory: Park:2012rh}.

Thus if the oscillatory fits really reflect inflationary features, there is a battery of consistency tests that the CMB temperature and polarization anisotropy must satisfy.

\section{Discussion}
\label{sec:discussion}

In this paper, we have extended and improved the modeling and analysis of sharp inflationary
steps for the Planck CMB power spectrum.  
We find that for the two parameters of the amplitude and frequency of the oscillations,
step models improve the fit by $\Delta \chi^2 = -11.4$ whereas additional
parameters such as the finite width of the step and sound speed of the inflaton 
marginally improve the fit to $\Delta \chi^2 = -14.0$ and $-15.2$ respectively.
In particular, sound speed effects for warp steps lower the quadrupole power by 
$\sim 20\%$.

We have shown that it is critical to jointly
fit step and cosmological parameters simultaneously.   If cosmological parameters, 
especially the amplitude and tilt, are held fixed then one would falsely infer that 
the oscillations must damp away at high multipole due to their excess average power. We have shown that on the contrary there
is only marginal preference for a finite damping scale.   The improvement in  modeling
to second order terms in the generalized slow roll approximation developed here is also required by the
increased precision of Planck at high multipoles but their omission would mainly bias
the cosmological parameters rather than degrade the fit itself.

Given that chance features in the noise can masquerade as oscillatory step features \cite{Meerburg:2013cla}, we have also provided a suite of consistency tests that can verify or falsify the primordial origin of
these improved fits.   The polarization power and cross spectra should reveal a matching and larger set of
oscillations modulated by an out of phase acoustic transfer.   The oscillations, if primordial, also 
provide an extra signal for CMB lensing reconstruction and squeezed bispectra
from the lensing-ISW correlation.   Finally, the primordial non-Gaussianity in equilateral
bispectrum configurations should also be observable.    These predictions may
soon be tested in the next release of Planck data.

\acknowledgments

We thank Peter Adshead, Aurelien Benoit-Levy, Douglas H. Rudd and Shi Chun Su for useful discussions.
WH was supported by the Kavli Institute for Cosmological
Physics at the University of Chicago through grants NSF
PHY-0114422 and NSF PHY-0551142 and an endowment
from the Kavli Foundation and its founder Fred Kavli.
VM and WH were additionally supported by U.S.\ Dept.\ of
Energy contract DE-FG02-90ER-40560,  WH by the David and Lucile
Packard Foundation and VM by the Brazilian Research Agency
CAPES Foundation and by U.S. Fulbright Organization.

%%%%%%%%%%%%%%%%%%%%%%%%%%%%%%%%%%%%%%%%%%%%%%%%%%%%%%%%%%%%%%%%%%%%%%%%%%%%%%%%%%%%%%%%%%%%%%%%%
\appendix
\section{Analytic Step Spectrum}
\label{app:analytic}

In this Appendix, we derive and test the analytic model for the power spectrum used
in the main paper, extending previous treatments \cite{Adshead:2011jq,Miranda:2012rm} for large, sharp steps in the warp and potential in the DBI context.  These necessitate second order corrections to achieve the precision required for the Planck data.   We begin with a brief review of the generalized slow roll (GSR)
approximation \cite{Stewart:2001cd,Hu:2011vr}
for large power spectrum features in \S \ref{sec:GSR} \cite{Dvorkin:2009ne}.  In \S \ref{sec:stepsources}, we use exact energy conservation and the
attractor solutions before and after the step to derive the general analytic model
for the power spectrum in the GSR approximation.   In \S \ref{sec:warpsteps}, \ref{sec:potentialsteps} we apply this
model to steps in the warp and potential.

%%%%%%%%%%%%%%%%%%%%%%%%%%%%%%%%%%%%%%%%%%%%%%%%%%%%%%%%%%%%%%%%%%%

\subsection{Generalized Slow Roll}
\label{sec:GSR}

	In a general $P(X,\phi)$ model for inflation, the comoving curvature power spectrum,
\begin{align}\label{eqn:power_spectrum}
	\Delta_{\R}^2 \equiv \frac{k^3 P_{\R}}{2 \pi^2} %        
	= \lim_{ks \to 0} \left| \frac{k s y}{f} \right|^2 ,
\end{align}
	is evaluated by solving the field or modefunction equation in spatially flat gauge \cite{Garriga:1999vw,Hu:2011vr}
\begin{align}\label{eqn:yeqn}
	\frac{d^2y}{ds^2} + \left(k^2 - \frac{2}{s^2} \right) y = \left(\frac{f'' - 3 f'}{f}\right)\frac{y}{s^2}.
\end{align}
Here deviations from de Sitter space are characterized by
\begin{align}\label{eqn:fdef}
	f^2 & = 8 \pi^2 \frac{\ep \cs}{H^2} \esq \frac{a H s}{\cs} \dir^2,
\end{align}
where
\begin{equation} \label{eqn:acceleration}
\epsilon_H=-\frac{d\ln H}{dN}
\end{equation}
with $N$ as efolds with $N=0$ as the end of inflation,
$c_s$ denotes the sound speed of field fluctuations, and 
\begin{align}\label{eqn:def_sound_horizon}
	s(N) = \int_{N}^{0} d \tilde{N} {c_s \over aH}
\end{align}
denotes the sound horizon.
Here and throughout $' \equiv d/d\ln s$. 

Eq.~(\ref{eqn:yeqn}) can be formally solved with the Green function technique by taking its 
right hand side as an external source given by the de Sitter mode function with
Bunch-Davies initial conditions
\begin{equation}
y \approx y_0 = \left( 1+ \frac{i}{k s} \right)e^{i k s} \,,
\end{equation}
and iteratively improving the solution for the presence of the deviations introduced by $f$.
Including the leading second order correction for large features, the power spectrum is given by 
  \cite{Choe:2004zg,Dvorkin:2009ne}
\begin{align} \label{eqn:GSRpower}
\ln \Delta_\curv^{2} &\approx  G(\ln s_{\rm min}) + \int_{s_{\rm min}}^\infty {d s\over s} W(ks) G'(\ln s)\\
&\quad + \ln \left[ 1+ I_1^2(k) \right], \nonumber
\end{align}
where the source function
\begin{equation}
G = - 2 \ln f    + {2 \over 3} (\ln f )'   ,
\end{equation}
and recall $W$ is given by Eq.~(\ref{eqn:powerwindow}).
The validity of the approximation relies on the deviations in the modefunctions or the curvature being small rather than the $G'$ deviations from slow roll.   It is 
monitored by the second order corrections
\begin{eqnarray}
I_1(k) &=& { 1\over \sqrt{2} } \int_0^\infty {d s \over s} G'(\ln s) X(ks),  
\end{eqnarray}
with $u=k s$ and $X$ given by Eq.~(\ref{eqn:powerwindow}).
The GSR approximation itself will begin to break down unless \cite{Dvorkin:2011ui}
\begin{equation}
I_1 \lesssim \frac{1}{\sqrt{2}} .
\label{eqn:I1criterion}
\end{equation}
.

Finally, $G'$ carries both the smooth tilt type deviations from scale invariance as well as 
any impact of sharp features.  For sharp  features it can be approximated by  \cite{Miranda:2012rm}
\begin{align}
	 G'  &\approx  (1-n_s) -\frac{1}{3}\sigma_2 + \frac{2}{3}\delta_2 -\frac{5}{3}\sigma_1 - 2\eta_H \nonumber\\& \quad+ \frac{8}{3}\left(\frac{aHs}{c_s}-1\right),
	\label{eqn:Gprimelead}
\end{align}
where the additional slow-roll parameters are defined by
\begin{align}\label{eqn:epsilon_def}
		 \eta_H &\equiv \epsilon_H - \frac{1}{2} \frac{d\ln\epsilon_H}{dN}, \nonumber\\
		\delta_2 &\equiv \epsilon_H \eta_H + \eta_H^2 - \frac{d\eta_H}{dN}, \nonumber\\
		\sigma_1 & \equiv \frac{d\ln c_s}{dN} ,\nonumber\\
		\sigma_2 & \equiv \frac{d\sigma_1}{dN} ,
\end{align}   
and we reabsorb the slow roll deviations in these parameters into the $(1-n_s)$ factor.
Thus modeling a sharp feature amounts to determining its impact on $c_s$ and $\epsilon_H$.

%%%%%%%%%%%%%%%%%%%%%%%%%%%%%%%%%%%%%%%%%%%%%%%%%%%%%%%%%%%%%%%%%%%%%%%%%%%%%%%%%%%%%%

\subsection{Step Sources}
\label{sec:stepsources}

	We consider sharp steps in the warp $T(\phi)$ and potential $V(\phi)$ of the DBI Lagrangian (\ref{eqn:DBI}) which generate 
	analogous changes in $c_s$ and $\epsilon_H$.  To keep our treatment general, we first parameterize the evolution of these quantities relying on energy conservation and the attractor 
	solution to define their functional form.
	In the following subsections we give the correspondence of this parameterization to
	specific step parameters.

	The energy density of the inflaton 
\begin{equation}
\rho = \left( {1\over c_s}-1\right)T + V
\end{equation}
is conserved as long as the inflaton rolls across the step in much less than an efold.
This conservation then gives the relationship between the sound speed 
before (``$b$"; $c_{sb}$) and immediately after (``$i$"; $c_{si}$) the step.   The acceleration equation 
\begin{eqnarray}
\epsilon_H &=& \frac{3}{2}\frac{\rho+p}{\rho} \approx \frac{3 T}{2 V} \left( \frac{1}{c_s}-c_s \right)
\label{eqn:epsV}
\end{eqnarray}
then gives the corresponding change in $\epsilon_H$.    After the step, the rolling of
the inflaton 
\begin{equation}
\phi_N^2 \equiv \left( \frac{d\phi}{dN} \right)^2 = 2 c_s \epsilon_H
\end{equation}
differs from the friction dominated attractor solution
\begin{equation}
\phi_N = -\frac{c_s}{3} \frac{V_\phi}{H^2} \approx - c_s \frac{V_\phi}{V}
\label{eqn:attractor}
\end{equation} 
to which it must decay on the expansion time scale or well after the inflaton has crossed the step (``$a$").   These relations hold for arbitrarily large steps so long as $\epsilon_H \ll 1$.

Together, they imply
that the functional form of $c_s$ and $\epsilon_H$ is generically given by \cite{Miranda:2012rm}
\begin{align}
\label{eqn:parammodel}
 \frac{c_s}{c_{sa}} =&1+ \frac{1-c_b}{2} F + \frac{c_i-1}{2} (F+2) e^{3(N_s-N)}, \\
\frac{\epsilon_{H}}{\epsilon_{Ha}} =&1+ \frac{1-e_b}{2} F + \frac{e_i-1}{2} (F+2) e^{3(N_s-N)},\nonumber
\end{align}
where for convenience we have scaled the quantities to their values on the attractor after the step
\begin{eqnarray}
c_b &  =  &\frac{c_{sb}}{c_{sa}},  \qquad\ c_i  =  \frac{c_{si}}{c_{sa}} , \nonumber\\
e_b &=& \frac{\epsilon_{Hb}}{\epsilon_{Ha}}, \qquad
e_i = \frac{\epsilon_{Hi}}{\epsilon_{Ha}} .
\label{eqn:csepsHmodel}
\end{eqnarray}
Here $F$ represents a step of infinitesimal width at $N=N_s$ normalized to $-2$ before the step and
$0$ after.   We discuss the impact of the finite width below.

Following Ref.~\cite{Miranda:2012rm}, it is straightforward to derive the source function $G'$
in the approximation that changes to $c_s$ and $\epsilon_H$ are small by taking their
derivatives and integrals to form the quantities in Eq.~(\ref{eqn:Gprimelead}).    Note that this limit does not necessarily require the steps in the warp itself to be small.    In the limit of a large warp factor $\phi_N^2/T \ll 1$,  the sound speed approaches unity regardless of the form of $T$ and hence the change in the sound speed are small even for a large fractional change in $T$.

Integrals over $G'$ are then simply evaluated
by recalling that $dF/d\ln s$ is a delta function of amplitude 2.    
The result of integrating the source by parts is
\begin{align} \label{eqn:GSR0}
	\ln \Delta_{\mathcal{R}}^2 \approx & 
	 \ln A_s \left(\frac{k}{k_0}\right)^{n_s-1}  +
	C_1 W (k s_s) + C_2 W' (k s_s) \nonumber \\ & + C_3 Y (k s_s) ,
\end{align}	
where
\begin{eqnarray}
C_1 &=& -(c_b-1)-(e_b-1), \nonumber\\
C_2 &=& -\frac{1}{3} (c_i-c_b) + \frac{1}{3} (e_i-e_b), \nonumber\\
C_3 &=& (1-c_b) + \frac{1}{4} (c_i-1),
\label{eqn:Csmall}
\end{eqnarray}
for the leading order GSR contribution in Eq.~(\ref{eqn:GSRpower}) which, once corrected for the finite width of the step below, we shall call GSR0.   Here we have replaced the parameter
$G(\ln s_{\rm min})$ in Eq.~(\ref{eqn:GSRpower}) with the power spectrum normalization $A_s$ at $k_0$.
Note that
\begin{equation}
Y(x)\equiv  -\frac{8}{3} x  \int d\ln \tilde x \frac{W'(\tilde x)}{\tilde x},
\end{equation}
which is given in closed form in Eq.~(\ref{eqn:powerwindow}).

Given that
\begin{align}
\lim_{x\ll 1} W(x) & = 1, \quad 
\lim_{x\gg1} W(x) = 0, 
\nonumber\\
\lim_{x\ll 1} W'(x) & =0, 
\quad \lim_{x\gg 1} W'(x) = -3\cos(2x),
\nonumber\\
\lim_{x\ll 1} Y(x) & =0, 
\quad \lim_{x\gg 1} Y(x) =0,
\end{align}
we can further interpret the meaning of the $C_i$ coefficients.    $C_1$  represents a step in the power spectrum and its amplitude is
determined by the fact that the inflaton is on the attractor solution before and well after the 
step.    $C_2$ provides a constant amplitude oscillation whose value is determined by the sharpest part of the feature: 
the fractional changes in $c_s$ and $\epsilon_H$ right at the step.    Finally $C_3$ 
modifies the shape of the first few oscillations due to the $a H s/c_s-1$
source.  

Likewise the first order corrections are given by
\begin{align}
\sqrt{2} I_1 = & \frac{\pi}{2}(1-n_s) + C_1 X (k s_s) + C_2 X' (k s_s)  
+ C_3 Z (k s_s)  ,
\label{eqn:I1analytic}
\end{align}	
where
\begin{align}
	Z(x) &=
-\frac{8}{3} x  \int d\ln \tilde x \frac{X'(\tilde x)}{\tilde x},
	\label{eqn:Zpowerwindow}
\end{align}
which is given in closed form in Eq.~(\ref{eqn:powerwindow}).

We call the analytic model with the $I_1$ correction GSR1.  Using Eq.~(\ref{eqn:I1criterion}), we thus expect the GSR expansion itself to be under control
for $k s_s \gg 1$ so long as
\begin{equation}
|C_2|<1/3.
\label{eqn:I1Cmax}
\end{equation}
At $k s_s \sim 1$, the exact requirement is a model dependent restriction on a
combination of  $C_1$, $C_2$, $C_3$ but in the warp and potential step examples this gives
roughly the same criteria for the step height.

In principle this domain of validity includes fractional deviations in 
$c_s$ and $\epsilon_H$ that approach unity, including the region of interest for Planck.   However although the GSR expansion itself remains under control, Eq.~(\ref{eqn:Csmall}) is derived by assuming small fractional deviations and requires correction.   Just as we extended the validity of the step approximation to nonlinearities in the step amplitude above, we can also approximately correct for weak nonlinearity in the slow roll parameters by rescaling the $C_i$ coefficients.
Here we extend and generalize the approach of  Ref.~\cite{Adshead:2011jq} and 
\cite{Miranda:2012rm} for arbitrary sound speeds.

   The $C_1$ amplitude gives the step in power and hence the slow roll attractor $\Delta_\curv^2 \propto (c_s \epsilon_H)^{-1}$ 
determines it  as
\begin{equation}
C_1 = -\ln c_b e_b.
\label{eqn:C1}
\end{equation}
The changes in $c_s$ and $\epsilon_H$ at the step are already determined nonlinearly and
so the only further correction to $C_2$ comes from the conversion to slow-roll parameters,
e.g. $\sigma_1  = c_s^{-1}dc_s/dN$.   Following Ref.~\cite{Adshead:2011jq}, 
we evaluate $c_s$ and $\epsilon_H$ at the midpoint of the step and hence
\begin{equation}
C_2 = -\frac{2}{3} \frac{c_i-c_b}{c_i + c_b}  + \frac{2}{3} \frac{e_i-e_b}{e_i+e_b}.
\label{eqn:C2}
\end{equation}
Finally for $C_3$, while there is no direct nonlinear constraint to determine its amplitude, by
also renormalizing to the midpoint of the step we approximately preserve the relative
relationship between the coefficients that determines the shape of their combined 
contributions. This is especially important for warp steps where
cancellations between $C_1$ and $C_3$ occur around the first oscillation.   Thus, we take
\begin{equation}
C_3 = 2\frac{(1-c_b) +(c_i-1)/4}{c_i+c_b}\,.
\label{eqn:C3}
\end{equation}

Finally, we can account for the finite width of the step.  If we replace the step function with
a $\tanh$ function 
\begin{align}
	F(\phi)= \tanh\Big(\frac{\phi-\phi_s}{d}\Big)-1,
\end{align}
the integrals over $G'$ will not contribute if the windows in Eq.~(\ref{eqn:powerwindow})  oscillate many times over
the width of the step $k \gg s_s  d/\phi_{N}$.    This causes a damping such that the
$C_i$ coefficients in Eqs.~(\ref{eqn:GSR0}) and (\ref{eqn:I1analytic}) are replaced by \cite{Adshead:2011jq}
\begin{equation}
C_i \rightarrow C_i  \mathcal{D}\Big(\frac{k s_s}{x_d}\Big),
\end{equation}
where the damping function
\begin{align}
	\mathcal{D}(y) = \frac{y}{\sinh(y)},
\end{align}
with the damping scale
	\begin{align}
	x_d = \frac{1}{\pi d} \frac{d\phi}{d\ln s}.
\end{align}
The derivation of Eq.~(\ref{eqn:ps_analytical_form}) for the functional form for the analytic model of the step power spectrum is thus complete.    We now turn to
specific forms for warp and potential steps.

\begin{figure}[t]
\psfig{file=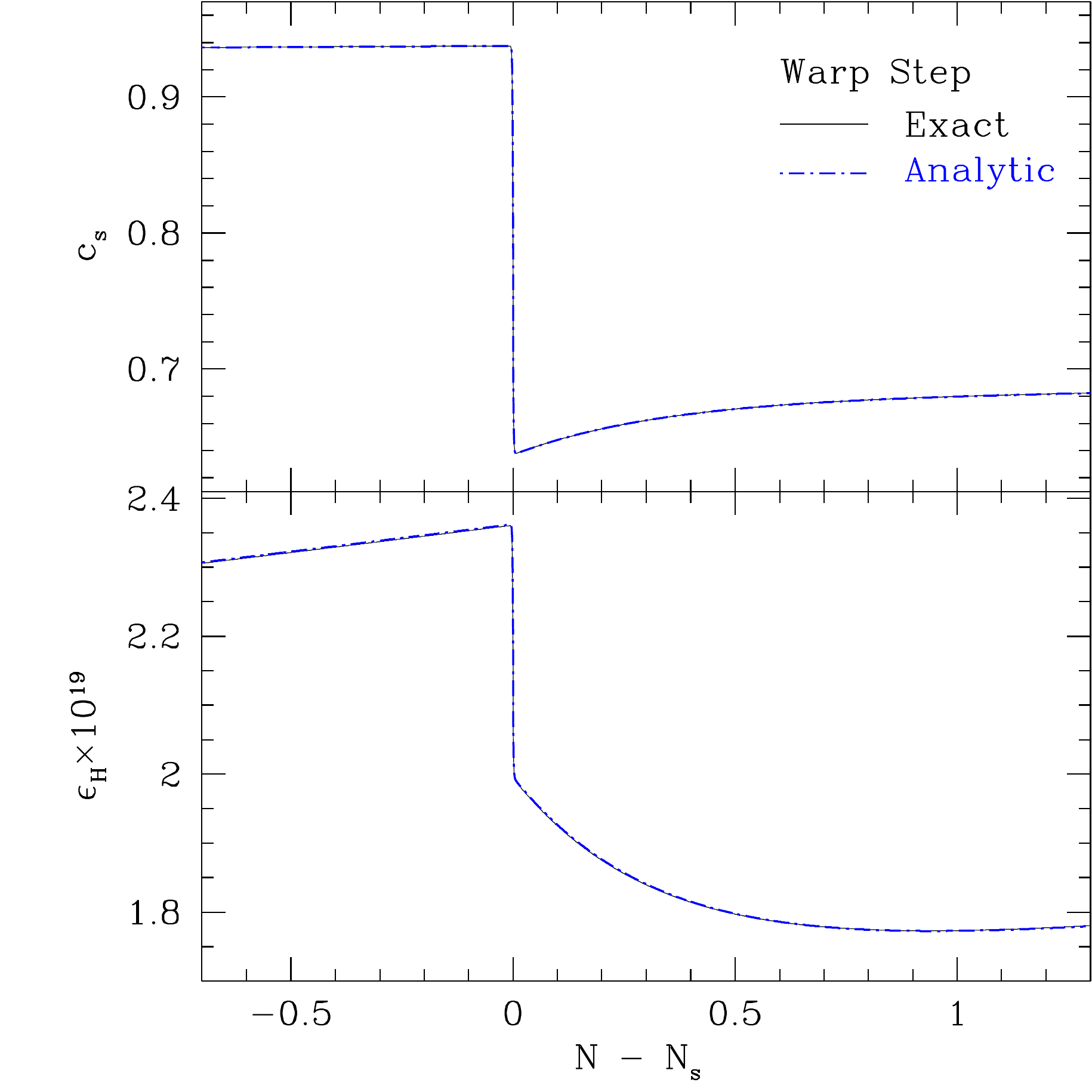, width=3.25in}
\caption{
Evolution of $c_s$ (upper) and $\epsilon_H$ (lower) across a warp 
step.   
The step in both parameters and  transient behavior right after the step is modeled
by  Eq.~(\ref{eqn:parammodel})  and (\ref{eqn:attractorcorrection}) 
 to excellent approximation. Model parameter choices are given in Appendix \ref{sec:accuracy}. 
}
\label{fig:warpevolution}
\end{figure}

\subsection{Warp Steps}
\label{sec:warpsteps}

For steps in the warp $T$, we have before and after the step
\begin{eqnarray}
T_b & =& T_a (1- 2 b_T),  \nonumber\\
V_b &= & V_a,
\end{eqnarray}
and we can then use energy conservation and the attractor solution to give the
relevant $c$ and $e$ parameters of the model in Eq.~(\ref{eqn:csepsHmodel}).  Our
convention is to quote these in terms of $b_T$ and the sound speed on the attractor after the step $c_{sa}$.
The attractor solution tells us that 
\begin{equation}
c_b = \sqrt{\frac{1-2b_T}{1-2b_T c_{sa}^2}} ,
\end{equation}
and 
using Eq.~(\ref{eqn:epsV}) for $\epsilon_H$, we obtain
\begin{equation}
e_b = c_b.
\end{equation}

Now let us consider the sharp changes immediately after the step.  Energy conservation tells us
\begin{equation}
\frac{1}{c_{si}} -1  = \left( \frac{1}{c_{sb}}-1 \right) (1-2 b_T),
\end{equation}
or 
\begin{eqnarray}
c_i &=& \frac{c_b }{1- 2b_T(1-c_{sa} c_b)} ,
%= f_b + \frac{2b(1-c_{sa}^2)}{(1- 2 b c_{sa}^2)(c_{sa}+f_b)}
\end{eqnarray}
and with Eq.~(\ref{eqn:epsV})
\begin{equation}
e_i = \frac{1- c_{sa}^2 c_i^2}{c_i(1-c_{sa}^2)}.
\end{equation}
We show an example of the evolution of $c_s$ and $\epsilon_H$ in Fig.~\ref{fig:warpevolution}.  {Since the analytic model only captures the evolution of the
parameters around the step and not the evolution on the slow roll attractor, we plot
\begin{equation}
 c_s(N-N_s)= {c_s^{\rm an}(N-N_s)} \frac{c_s^{\rm at}(N-N_s)}{c_s^{\rm at}(0^\pm)},
 \label{eqn:attractorcorrection}
\end{equation}
where $c_s^{\rm an}$ is the analytic model of Eq.~(\ref{eqn:parammodel}), 
 $c_s^{\rm at}$ is the attractor on either side of the step and $c_s^{\rm at}(0^\pm)$ is
evaluated approaching the step from either side with $c_{sa}=c_s^{\rm at}(0^+)$ approached
from the side after the step.     We likewise account for the slow roll evolution of $\epsilon_H$. In practice, rather than iterating the attractor solution of Eq.~(\ref{eqn:attractor}) in the
equations of motion to 
the required accuracy we numerically solve the equivalent smooth model before
and after the step to determine $c_s^{\rm at}$. }

We can now use the general description of Eqs.~(\ref{eqn:C1}), (\ref{eqn:C2}), (\ref{eqn:C3})
to give the $C_i$ coefficients of the analytic power spectrum form.
Note that in the small step limit,
\begin{eqnarray}
\lim_{b_T \rightarrow 0} c_b &=& 1-(1-c_{sa}^2) b_T, \nonumber\\
\lim_{b_T \rightarrow 0} c_i &=& 1+(1-c_{sa})^2 b_T, \nonumber\\
\lim_{b_T \rightarrow 0} e_i &=& 1-\frac{(1-c_{sa})(1+ c_{sa}^2) b_T}{1+ c_{sa}} , 
\end{eqnarray}
and so
\begin{eqnarray}
\lim_{b_T \rightarrow 0} C_1 &=& 2(1-c_{sa}^2)b_T,\nonumber\\
\lim_{b_T \rightarrow 0} C_2 &=& -\frac{2}{3}\frac{1-c_{sa}}{1+c_{sa}}b_T,\nonumber\\
\lim_{b_T \rightarrow 0} C_3 &=& \frac{1}{4} (5-2 c_{sa}-3 c_{sa}^2) b_T,
\end{eqnarray}
in agreement with Ref.~\cite{Miranda:2012rm}.  Our generalized expression lets us explore
the $b_T \rightarrow -\infty$ limit
\begin{eqnarray}
\lim_{b_T \rightarrow -\infty} c_b &=& \frac{1}{c_{sa}}, \nonumber\\
\lim_{b_T \rightarrow -\infty} c_i &=& \frac{2 c_{sa}}{1+ c_{sa}^2}, \nonumber\\
\lim_{b_T \rightarrow -\infty} e_i &=& \frac{1}{2 c_{sa}} \frac{1+ 3 c_{sa}^2}{ 1+ c_{sa}^2}.
\end{eqnarray}
Note that for finite $c_{sa}$ these limits are all finite and so the maximal
$C_i$ amplitudes are also bounded
\begin{eqnarray}
\lim_{b_T \rightarrow -\infty} C_1 &=& 2\ln c_{sa} ,\nonumber\\
\lim_{b_T \rightarrow -\infty} C_2 &=&4\frac{1- c_{sa}^4}{9+42 c_{sa}^2+ 45 c_{sa}^4},\nonumber\\
\lim_{b_T \rightarrow -\infty} C_3 &=&-\frac{1}{2} \frac{4-3 c_{sa}+ 2 c_{sa}^2-3 c_{sa}^3}{1+3 c_{sa}^2}.
\end{eqnarray}
Thus for a fixed observed oscillation amplitude $C_2>0$ there is always a maximum $c_s$ for
which a warp step cannot explain the data
\begin{equation} \label{eqn::csmaxwarp}
c_{sa}^2 \Big|_{\rm max} = -\frac{21 C_2}{4+ 45 C_2} + \frac{2\sqrt{4+36 C_2+ 9 C_2^2}}{4 + 45 C_2}.
\end{equation}  
For example, if $C_2 = 1/15$, $c_{sa}^2|_{\rm max} \approx 0.724$.

\begin{figure}[t]
\psfig{file=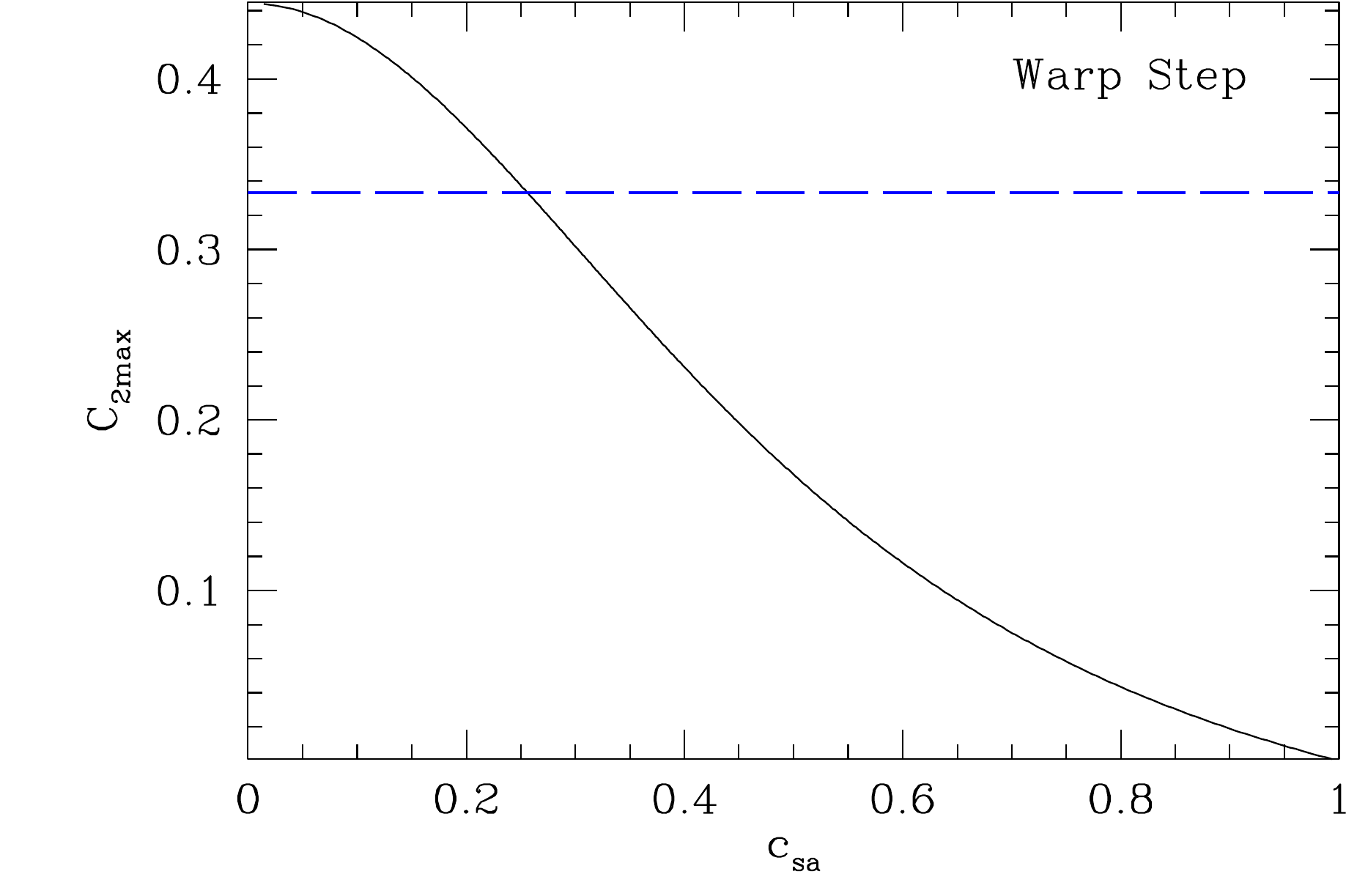, width=3.25in}
\caption{ Maximum oscillation amplitude $C_2$ for a warp step as a function of 
$c_{sa}$, the sound speed after the step.   The blue dashed line corresponds to $C_2=1/3$  where the
GSR approximation breaks down in the oscillatory regime.  Note that for $c_s \gtrsim 1/4$
the oscillation amplitude is limited by physicality rather than the GSR approximation.}
\label{fig:SR_variations}
\end{figure} \label{figure:csmax_warp}

\begin{figure}[t]
\psfig{file=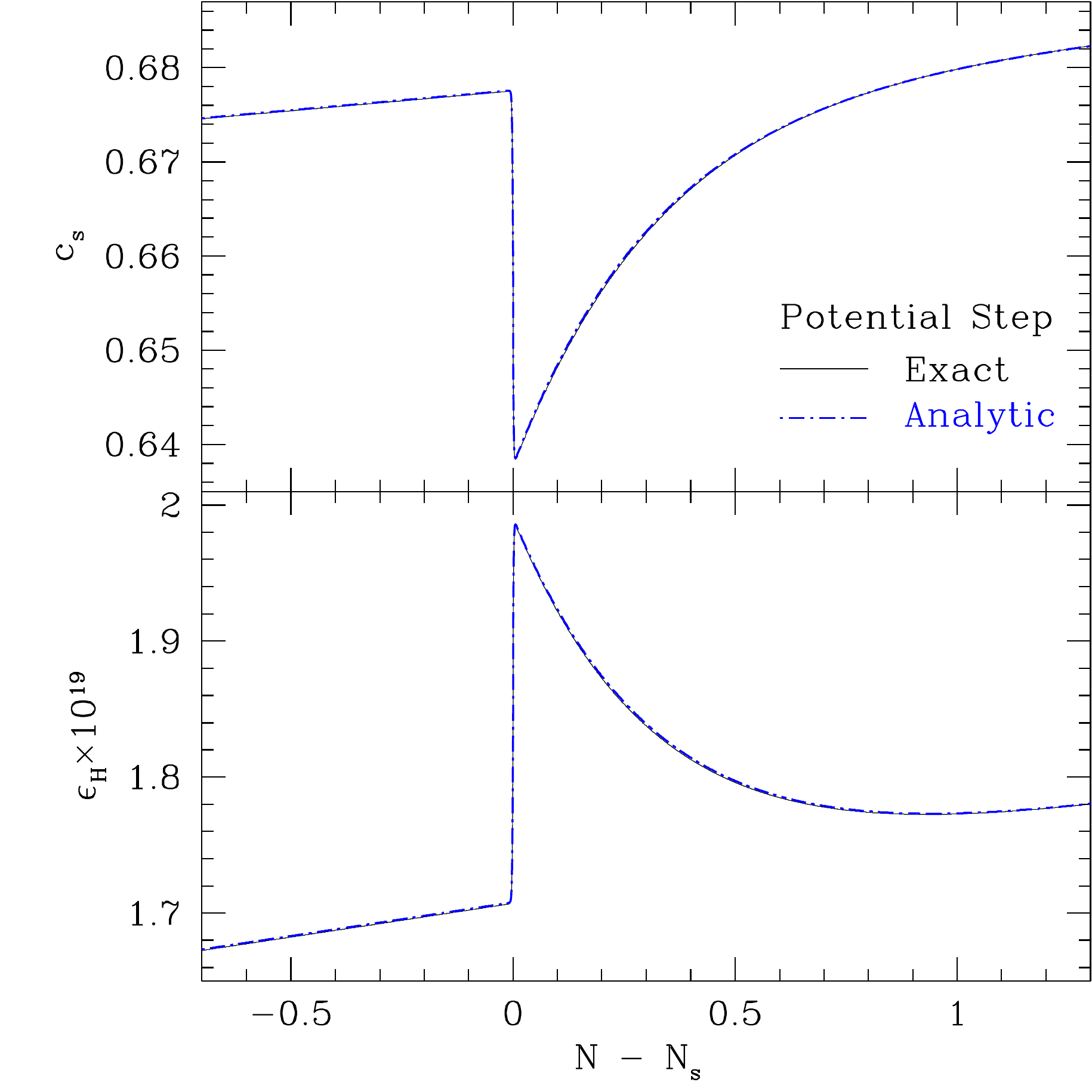, width=3.25in}
\caption{Evolution of $c_s$ (upper) and $\epsilon_H$ (lower) across a potential 
step.  The transient change in both parameters is modeled by  Eq.~(\ref{eqn:parammodel}) and 
(\ref{eqn:attractorcorrection}) to excellent approximation.
Model parameter choices are given in Appendix \ref{sec:accuracy}.  }
\label{fig:potentialevolution}
\end{figure}

%%%%%%%%%%%%%%%%%%%%%%%%%%%%%%%%%%%%%%%%%%%%%%%%%%%%%%%%%%%%%%%%%%%%%%%%%%%%%%%%%%%%%%%%

\subsection{Potential Steps}	
\label{sec:potentialsteps}

For potential steps
\begin{eqnarray}
V_b &=& V_a(1-2b_V), \nonumber\\
T_b &=& T_a.
\end{eqnarray}
 The attractor solution says that to leading
order in $\epsilon_{H}$, there is no net change in $c_s$ or $\epsilon_H$ only a transient
deviation at the step.   
 Thus
$\epsilon_{Hb}=\epsilon_{Ha}$ and $c_{sa}=c_{sb}$ ($c_b=e_b=1$) or
\begin{equation}
C_1=0.
\end{equation}
We can also use the attractor to eliminate $T/V_a$ using
\begin{equation}
\epsilon_{Ha} = \frac{3}{2} \frac{T}{V_a} \left( \frac{1}{c_{sa}}-c_{sa} \right).
\end{equation}

Energy conservation then gives the transient change as
\begin{equation}
c_i =1 -\frac{ 3 b_V (1- c_{sa}^2)}{3 b_V (1- c_{sa}^2) - \epsilon_{Ha}},
\end{equation}
and using Eq.~(\ref{eqn:epsV}) 
\begin{equation}
e_i=1
-\frac{ 3 b_V [-3b_V(1- c_{sa}^2) + (1+c_{sa}^2) \epsilon_{Ha}]}{\epsilon_{Ha} [-3 b_V(1- c_{sa}^2)+ \epsilon_{Ha} ] }.
\end{equation}
The general description of Eqs.~(\ref{eqn:C1}), (\ref{eqn:C2}), (\ref{eqn:C3})
then gives the $C_i$ coefficients of the analytic power spectrum form.
We show an example of the evolution of $c_s$ and $\epsilon_H$ in Fig.~\ref{fig:potentialevolution}. {Again, to capture the slow roll evolution of the smooth model, we
plot the analytic model corrected as in Eq.~(\ref{eqn:attractorcorrection}).}

In the limit of a small potential step
\begin{eqnarray}
\lim_{b_V\rightarrow 0}c_i &=& 1+ 3 \frac{1-c_{sa}^2}{\epsilon_{Ha}} b_V, \nonumber\\
\lim_{b_V\rightarrow 0}e_i &=& 1- 3 \frac{1+c_{sa}^2}{\epsilon_{Ha}} b_V,
\end{eqnarray}
and so
\begin{eqnarray}
\lim_{b_V\rightarrow 0}C_2 &=& - \frac{2}{\epsilon_{Ha}} b_V, \nonumber\\
 \lim_{b_V\rightarrow 0}C_3 &=& \frac{3}{4}\frac{1-c_{sa}^2}{\epsilon_{Ha}} b_V,
 \end{eqnarray}
which generalizes the results of Ref.~\cite{Adshead:2011jq} to arbitrary sound speed.
The sound speed experiences a transient dip for downward steps $b_V<0$.   

Note that in the opposite limit
\begin{eqnarray}
\lim_{b_V\rightarrow -\infty} c_i &=& \frac{1}{3b_V}\frac{\epsilon_{Ha}}{1-c_{sa}^2} , \nonumber\\
\lim_{b_V\rightarrow -\infty} e_i &=& \frac{3b_V}{\epsilon_{Ha}},
\end{eqnarray}
and so
\begin{eqnarray}
\lim_{b_V\rightarrow -\infty} C_2 &=& \frac{4}{3}, \nonumber\\
\lim_{b_V\rightarrow -\infty} C_3 &=& -\frac{1}{2} .
\end{eqnarray}
Since these amplitudes are beyond the limits of the GSR approximation itself 
according to Eq.~(\ref{eqn:I1Cmax}), there is effectively no relevant bound on the oscillation amplitude set by 
energy conservation and the attractor solution unlike the warp step case.  Likewise,
for a given $0<C_2 \ll 4/3$, there is no bound on the required sound speed.

\section{Power Spectrum Accuracy}
\label{sec:accuracy}

In this section, we test the accuracy of the leading order GSR0 approximation used in previous analyses \cite{Adshead:2011jq}
 and the first order GSR1 corrections discussed in Appendix \ref{app:analytic} 
against an exact computation of the power spectrum from the DBI Lagrangian of Eq.~(\ref{eqn:DBI}).  Although  GSR0 was previously demonstrated to be sufficiently accurate for WMAP  data \cite{Adshead:2011jq,Miranda:2012rm}, we show here that the increase in precision
to the $10^{-3}$ level in Planck requires second order corrections.

The exact computation of the power spectrum follows from solving Eq.~(\ref{eqn:yeqn})
for a DBI step model that is parameterized by $\{V_0, \beta, \lambda_b,  \phi_{\text{end}}\}$,
defining the broadband amplitude and slope of the power spectrum,
and the step parameters $\{ \phi_s, b_T, b_V, d \}$ defining the step position, height parameters, and width (see \cite{Miranda:2012rm} for computational details).   For testing purposes, we choose
\begin{eqnarray}
V_0 &=& 7.1038 \times 10^{-26} ,\nonumber\\
\beta &=& 5.5895 \times 10^{-2} ,\nonumber\\
\lambda_b &=& 2.1771  \times 10^{14} ,\nonumber\\
\phi_{\text{end}} &=&  8.2506 \times 10^{-8},
\label{eqn:DBIparam}
\end{eqnarray}
and 
\begin{eqnarray}
\phi_s &=& 3.8311 \times 10^{-8} ,\nonumber\\
d &=& 9.3835 \times 10^{-13}.
\end{eqnarray}

For the warp step, we choose 
\begin{eqnarray}
b_T &=& -3.364, \nonumber\\\
b_V &=& 0, \quad ({\rm warp})
\end{eqnarray}
and for the  potential step 
\begin{eqnarray}
b_T &=& 0 ,\nonumber\\\
b_V &=& -6.543  \times 10^{-21}, \quad ({\rm potential}).
\end{eqnarray}
These parameters are in fact chosen to be close to the Planck maximum likelihood 
solution for the amplitude and frequency of warp step oscillations  by inverting the steps in this test.  Notice that in that case, $|b_T|$ the
fractional change in the warp $T$ exceeds unity.  The width $d$ is set so that damping occurs in the $\ell\sim 10^3$ region that Planck is most sensitive to so as to yield the most
stringent test of accuracy.  The cosmological parameters for the test are given in
Tab.~\ref{table:smooth_parameters} and coincide with the best fit model without a step.

%\begin{table}[t] \centering
%\def\arraystretch{1.40}
%\begin{tabular}{| c | c | }
%\hline
% $100\theta_A$ & $1.04136$   \\  
% $10\Omega_c h^2$ & $1.2035$ \\
% $100\Omega_b h^2$ & $2.2053$  \\
%  $100\tau$ & $8.952 $ \\ \hline   
%\end{tabular}
%\caption {Cosmological Parameters set on accuracy tests}  
% \label{table:test_accuracy_parameters} 
%\end{table}

The analytic models are specified by the conversion of the fundamental parameters into 
the 
amplitude parameters $\{ C_1, C_2, C_3 \}$,
the sound horizon at the step $s_s$, the effective number of oscillations before damping
$x_d$, as well as the broadband amplitude and tilt parameters $A_s$ and $n_s$.

Given a solution to the background equations without the step,
 we set the parameters
 \begin{eqnarray}
 c_{sa}&=& 0.67, \nonumber\\
 \epsilon_{Ha}&=& 1.70 \times 10^{-19},
 \end{eqnarray}
 according to their values at $N=N_s$. 
The $C_i$ amplitude parameters are then determined by Eqs.~(\ref{eqn:C1}-(\ref{eqn:C3})
such that

 \begin{eqnarray}
C_1 &=& -0.65, \nonumber\\
C_2 &=&  0.071, \nonumber\\
C_3 &=&  -0.34,   \qquad ({\rm warp})
\end{eqnarray}
and 
 \begin{eqnarray}
C_1 &=& 0, \nonumber\\
C_2 &=& 0.071, \nonumber\\
C_3 &=&  -0.015	,			\qquad ({\rm potential}).
\end{eqnarray}

Next, the physical scale associated with the step has to be set very precisely in order not
to have a phase error after many oscillations.  We follow Ref.~\cite{Miranda:2012rm}
in defining it numerically to be the sound horizon at which the deviation in the
GSR source function due to the step is appropriately centered to a small fraction of the step width
\begin{align}
	G'(\ln s_s, b_{T,V})-G'(\ln s_s, 0)=0.
\end{align}
Using this definition we obtain 
\begin{eqnarray}
s_s =
\begin{cases}
3699 \, {\rm Mpc} & ({\rm warp}) \\
3708 \, {\rm Mpc} & ({\rm potential})
\end{cases}.
\end{eqnarray}
Note that although the step is at the same position in field space in both cases, the sound
horizon differs slightly due to the change in $c_s$.

\begin{figure*}[t]
\psfig{file=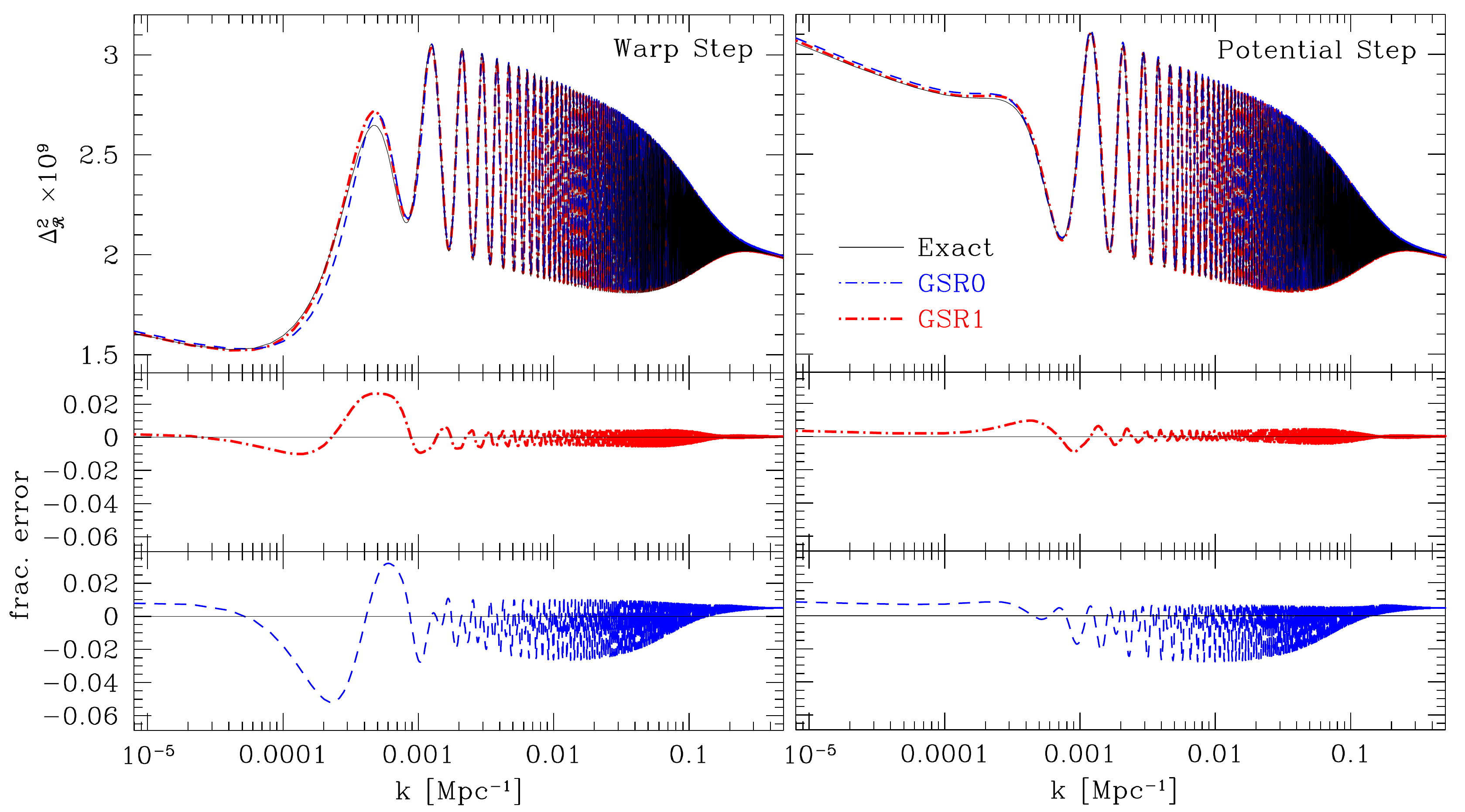, width=5.5in}
\caption{GSR approximations vs exact solution for the curvature power spectrum (top panel) and the fractional error of the GSR0 and GSR1 analytical solutions (bottom panel). The models are warp step (left) and potential step (right).}
\label{fig:test_analytics}
\end{figure*}

\begin{figure*}[t]
\psfig{file=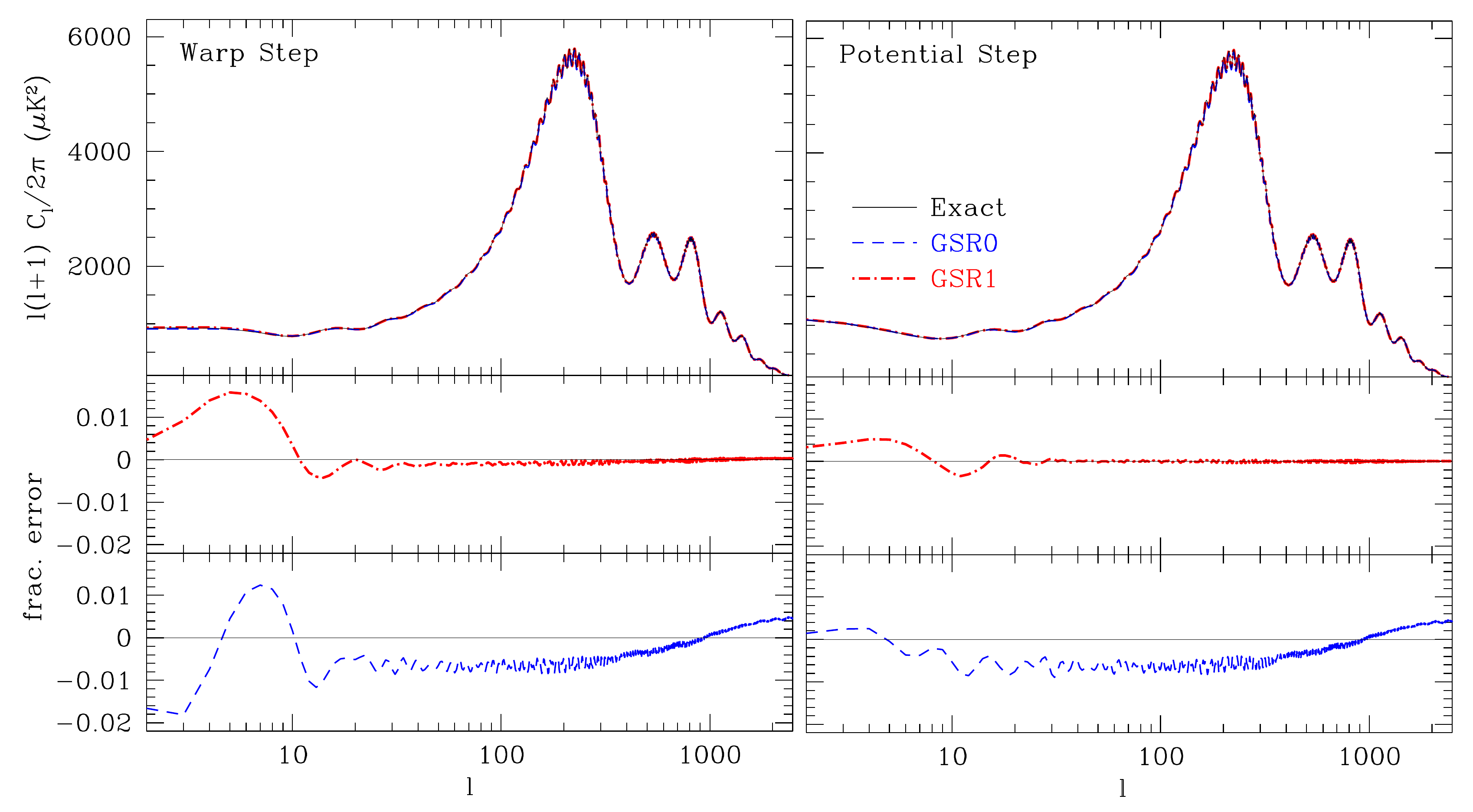, width=5.5in}
\caption{GSR approximations vs exact solution for the temperature power spectrum (top panel) and the fractional error of the GSR0 and GSR1 analytical solutions (bottom panel). The models are warp step (left) and potential step (right).}
\label{fig:test_analytics_Cl}
\end{figure*}

For the damping parameter, we likewise  convert the field width $d$ to 
a physical width $s_s/x_d$ with the numerical solution for $\phi(\ln s)$ through
\begin{align}
	\frac{d\phi}{d\ln s}\frac{1}{\pi d}\Bigg|_{s=s_s} = x_d.
\end{align}
For the test cases, we obtain 
\begin{eqnarray}
x_d =
\begin{cases}  170.0 & ({\rm warp}) \nonumber\\
 169.9 & ({\rm potential})
 \end{cases}.
\end{eqnarray}

Finally there are the broadband power parameters $n_s$ and $A_s$.  For the
tilt parameter, which is slowly varying and essentially independent of the step,
we take the slope at $k=k_0=0.08$ Mpc$^{-1}$ of the model with $b_{T,V}=0$.   We have chosen parameters in Eq.~(\ref{eqn:DBIparam}) so that $n_s$ coincides with
the value given in 
Tab.~\ref{table:smooth_parameters}. On the other hand, the effective amplitude  $A_s$ depends on the presence of the
step as well as the order of the GSR approximation used.   In Eq.~(\ref{eqn:GSR0}),
 the broadband  power gains a contribution from the average of the oscillations
 \begin{eqnarray} \label{eqn:average_oscillation}
 \langle e^{C_1 W + C_2 W' + C_3 Y} \rangle &\approx& {\cal I}_0\left[ 3C_2  {\cal D}\left( \frac{k s_s}{x_d} \right)\right] \\
 &\approx & 1 + \left[ \frac{3}{2} C_2 {\cal D}\left( \frac{k s_s}{x_d} \right)\right]^2+{\cal O}(C_2^4),\nonumber
 \end{eqnarray}
 where ${\cal I}_0$ here is the modified Bessel function, not to be confused with the GSR integral $I_0$.  This non-zero average
  is the fundamental reason why cosmological parameters must be varied jointly with the step parameters
 when analyzing the Planck data. In the first order correction Eq.~(\ref{eqn:GSRpower}), there is the analogous
 averaging effect 
 \begin{eqnarray}
 \langle I_1^2 \rangle \approx \frac{\pi^2}{8}(1-n_s)^2 + \left[ \frac{3}{2} C_2 {\cal D}\left( \frac{k s_s}{x_d} \right)\right]^2,
  \label{eqn:average_oscillation1}
 \end{eqnarray}
 which is also ${\cal O}(C_2^2)$ despite being higher order in the GSR approximation.
 Moreover, around the damping scale set by $x_d$ the broadband average of the 
 oscillation changes with $k$ in Eq.~(\ref{eqn:average_oscillation})-(\ref{eqn:average_oscillation1}) and is not purely an amplitude shift.   
% For models such as our test case where the damping falls in the region Planck constrains
 %the best, this causes an error in GSR0 that cannot be compensated by a change in $A_s$.
 Note that the error induced by this average term scales as $\delta C_l/C_l \propto C_2^2$ and so rapidly increases with the amplitude of the oscillations.
 
 Since the best choice for $A_s$ depends on both the method and the data set 
 considered, we choose $A_s$ as the amplitude which gives the best agreement between
 the exact computation and the given GSR computation
 for the Planck dataset.  We therefore use the Planck likelihood itself to define $A_s$
 for each method.   In order to remove the ambiguity caused by the exact model not
 possessing the maximum likelihood normalization, we in practice maximize both the Planck likelihood over $A_s$ to obtain $A_s'$ 
 and a rescaling of the amplitude of the exact  model by $R$ to obtain its best normalization.
 We then set $A_s= A_s'/R$ to remove the rescaling.

In Fig.~\ref{fig:test_analytics_Cl} we show the residual errors in the GSR0 and GSR1
after the normalization has been set in this way.   Note that the residuals $\delta C_l/C_l$ 
cross zero at $\ell \sim 10^3$, reflecting the pivot or best constrained portion of the Planck spectrum.
For scales much smaller than or much larger than the damping scale of the oscillation, 
the difference between GSR0, GSR1, and exact is nearly constant and can be absorbed
into the normalization.   However in our test case, which we have chosen to be the worst
case scenario, the damping falls exactly  at the pivot.   The result is that even with the
best fit normalization, GSR0 produces $\sim 1\%$ errors that pivot around $\ell \sim 10^3$.
While the error in GSR0 can mainly be absorbed by adjusting cosmological
parameters such as the tilt, they are large enough to bias such parameters non-negligibly. 
These residuals are reduced to the $\sim 0.1\%$ level with the GSR1 approximation.

More quantitatively, for these specific test cases
the residuals
produce a change in the Planck likelihood versus exact of
\begin{align} 
	\Delta\chi^2 =\begin{cases}
	 -8.6 & \text{GSR0}  \\
	 -0.97 & \text{GSR1} \end{cases} \quad ({\rm warp}),  
\end{align}	
and 
\begin{align}
	\Delta\chi^2 = \begin{cases}
	-7.4 & \text{GSR0}  \\
    -0.33 & \text{GSR1} \end{cases}   \quad ({\rm potential}).
\end{align}
Thus the GSR1 approximation is sufficiently accurate for the Planck analysis.  In fact, the 
$\chi^2$ errors would be even smaller at its global minimum.

The error in these approximations also depends on the step parameter model.  For reference if $x_d \rightarrow \infty$ the error in the GSR0 approximation becomes
$\Delta\chi^2=0.5$ for the warp step and $\Delta\chi^2=0.7$ for the potential step.   Here the error is
significantly lower since it takes the form of a constant amplitude rescaling which
can be absorbed into $A_s$.

In our examples the result of the normalization procedure is to set,
\begin{align}
	A_s  =
	\begin{cases} 
	 2.1432  \times 10^{-9}  & \text{GSR0}  \\
	2.1295 \times 10^{-9} &  \text{GSR1}  
	\end{cases}\quad ({\rm warp}), 
	\label{eqn:Aswarp}
\end{align}
for the warp step and
\begin{align} 
	A_s  =
	\begin{cases} 
	 2.1425 \times 10^{-9} & \text{GSR0}  \\
	 2.1288 \times 10^{-9} &  \text{GSR1} 
	\end{cases} \quad ({\rm potential}),  
	\label{eqn:Aspot}
\end{align}
for the potential step.   In the absence of a step ($b_T=b_V=0$), the same
procedure would yield $A_s=2.1554 \times 10^{-9}$ which is $1\%$ different from the GSR1 
value.  These changes reflect the broadband power
introduced by the oscillations in each case. 
 Given the $0.1\%$ precision of the Planck data  these differences are
significant and $A_s$ cannot be held fixed when fitting to step models.

For the minimization procedure in Appendix \ref{sec:minimize}, it is nonetheless useful to have
an approximate prescription for the renormalization of $A_s$ in the presence
of oscillations.   Given the average of the oscillatory pieces in Eq.~(\ref{eqn:average_oscillation}-\ref{eqn:average_oscillation1}), the normalization parameter
that the data should hold approximately fixed is 
\begin{align}
 \tilde A_s = A_s e^{-2\tau}(1+ \bar O),
 \label{eqn:Astilde}
\end{align}
where $\bar O$ contains the average of the oscillatory pieces in
each approximation
\begin{align}
\bar O=
 \begin{cases}
\frac{9}{4} C_2^2 {\cal D}^2 \left( \frac{k_0 s_s}{x_d} \right) &  {\rm GSR0} \\
 \frac{\pi^2}{8}(1-n_s)^2+
\frac{9}{2} C_2^2 {\cal D}^2\left( \frac{k_0 s_s}{x_d} \right) &{\rm GSR1}
 \end{cases}
  \,.
\end{align}
Here the $e^{-2\tau}$ factor accounts for the change in the heights of the acoustic peaks
due to anisotropy suppression by scattering during reionization.  
In our test cases
\begin{align}
	\tilde A_s  =
	\begin{cases} 
	1.8000 \times 10^{-9}  \quad \text{GSR0}  \\
	1.7999 \times 10^{-9} \quad  \text{GSR1}  
	\end{cases}\quad ({\rm warp}), 
	\label{eqn:Astildewarp}
\end{align}
for the warp step,
\begin{align} 
	\tilde A_s  =
	\begin{cases} 
	 1.7993 \times 10^{-9} \quad \text{GSR0}  \\
	 1.7993 \times 10^{-9} \quad  \text{GSR1} 
	\end{cases} \quad ({\rm potential}),  
	\label{eqn:Astildepot}
\end{align}
for the potential step and $\tilde{A}_s=1.8021 \times 10^{-9}$ without a step.
Note that $\tilde A_s$ absorbs most of the changes in the $A_s$ normalization 
given in Eq.~(\ref{eqn:Aswarp}-\ref{eqn:Aspot}) from the presence
of the step.

\begin{table}[t] \centering
\def\arraystretch{1.40}
\begin{tabular}{| c | c | }
\hline  
 $10^9 \tilde{A}_s$ & $1.8027$  \\  
 $n_s$ & $0.9607$  \\
 $100\theta_A$ & $1.04144$  \\  
 $10\Omega_c h^2$ & $1.1995 $ \\
 $ 100\Omega_b h^2$ & $2.2039$ \\
 $100\tau$ & $8.952 $ \\\hline
 $H_0$ & $67.22$ \\
 $10^9 A_s$ & $2.1562$ \\
 $D_* (\text{Mpc})$ & $13893.1$ \\\hline
 $\chi^2$ & $9802.8$ \\   \hline
\end{tabular}
\caption { Best fit flat $\Lambda$CDM cosmological model without a step with 6 varied parameters (top) and derived parameters (bottom). This model provides
the baseline $\chi^2$ for the smooth model but its parameters require adjustment in the
presence of a step.   $\tilde A_s$ is an  effective normalization parameter defined in 
Eq.~(\ref{eqn:Astilde}). }  
 \label{table:smooth_parameters} 
\end{table} 

\begin{table}[t] \centering
\def\arraystretch{1.40}
\begin{tabular}{| c | c || c | c | }
\hline
  $\gamma^{\rm CIB}$ & $0.538$ & $A^{\rm PS}_{217} $ & $112.4$  \\  
  $r^{\rm PS}_{143\times217}$ & $0.906$  & $A^{\rm CIB}_{143} $ & $6.18$  \\
  $A^{\rm CIB}_{217}$ &  $27.5$ & $c_{100}$ & $1.000580$     \\  
  $A^{\rm tSZ}_{143}$ & $6.71$ & $c_{217}$ & $0.9963$   \\
  $\xi^{\rm tSZ-CIB}$ & $0.2$   & $\beta^1_1$ & $0.55$   \\ 
  $A^{\rm PS}_{100}$ & $152$ & $A^{\rm kSZ}$ & $3$   \\
  $A^{\rm PS}_{143}$ & $50.8$ &  $r^{\rm CIB}_{143\times217}$ & $0.365$     \\  \hline
\end{tabular}
\caption { Foreground model.  These parameters are jointly minimized with
those of Tab.~\ref{table:smooth_parameters} in the smooth model and held fixed for
the step analysis.} 
 \label{table:foregrounds} 
\end{table}

\section{Minimization}
\label{sec:minimize}

In this Appendix we provide details of the effective $\chi^2$ minimization for the various models presented in \S \ref{sec:planck}.   In each case, we use the MIGRAD variable metric minimizer from the CERN Minuit2 code
\cite{James:1975dr}. 

We begin with the  smooth
$\Lambda$CDM cosmology specified by the cosmological parameters
$\{\tilde A_s, n_s, \theta_A, \Omega_c h^2, \Omega_b h^2, \tau \}$, and 14 foregrounds parameters  defined in the Planck likelihood \cite{Planck:2013kta}. We include the Planck low-$\ell$ spectrum (Commander, $l<50$), the high-$\ell$ spectrum (CAMspec, $50<l<2500$) and WMAP9 polarization (lowlike) likelihoods in our analysis  \cite{Planck:2013kta, Bennett:2012fp}.   $\tilde A_s$ is
the effective normalization defined in Eq.~(\ref{eqn:Astilde}); in the absence of a step
$\tilde A_s=A_s e^{-2\tau}$.
In the standard $\Lambda$CDM model, the effective number and sum of the masses 
of neutrinos are held fixed to $N_{\text{eff}} = 3.046$ and $\sum m_{\nu} = 0.06 \text{eV}$
respectively with the helium fraction $Y_P = 0.2477$.   The best fit model is given in Tab.~\ref{table:smooth_parameters}
and \ref{table:foregrounds}
and its $\chi^2$ 
 is the baseline from which we quantify any improvements due to the step
parameters.   We fix the foreground
parameters to these values for the following analysis but have spot checked that
reoptimization over the foreground parameters does not substantially change our results.

\begin{table} [t]
\def\arraystretch{1.40}
\begin{tabular}{ | c | c | c | c | c |   }
\hline
 & \multicolumn{2}{c|}{GSR1} & \multicolumn{2}{c|}{GSR0}  \\\hline 
 $x_d$ & $105 $ &  $2000 $ & $105 $ &  $2000$  \\
  $10 \theta_s$ & $2.665$  & $2.667$ & $2.665$  & $2.666$ \\
  $10 A_c$ & $1.17$  &  $0.663$ & $1.11$  &  $0.707$ \\ 
  $10^9 \tilde{A_s}$ &  $1.8021$  & $1.8024$ & $1.8020$  & $1.8021$ \\
 $n_s$ & $0.9690$  &  $0.9608$ & $0.9640$  &  $0.9606$ \\
 $100\theta_A$ &  $1.04140$  &  $1.04145$   &  $1.04136$  &  $1.04140$ \\
 $10\Omega_c h^2$ &  $1.2091$ & $1.1995$ &  $1.2035$ & $1.1993$   \\ 
 $100 \Omega_b h^2$&   $2.1974$  &  $2.2039$ & $2.2053$  &  $2.2039$ \\ 
 $100\tau$ & $9.421 $ & $9.117$ & $9.361$ & $9.205$ \\ \hline  
 %  Derived parameters H0, As, C2, s_s
 $H_0$ & $66.82$ & $67.23$ & $67.07$ & $67.22$ \\
 $10^9 A_s$ & $2.1669$ & $2.1420$ & $2.1701$ & $2.1565$ \\
  $D_* (\text{Mpc})$ & $13874.7$ & $13893.2$ &  $13882.9$ & $13893.9$\\
   $s_s (\text{Mpc})$ & $3696.9$ & $3704.7$ & $3699.2$ & $3704.5$ \\ 
 $C_2$ & $0.075$ & $0.043$ & $0.071$ & $0.045$ \\\hline  
 $\Delta\chi^2$ & $-14.0$ & $-11.4$ & $-13.8$ & $-11.3$ \\\hline   
\end{tabular}
\caption {Best fit potential step model with $c_s=1$ showing the 9 parameters jointly varied (top) and derived parameters (bottom)
using the GSR1 approximation of the main paper versus the less accurate GSR0 approximation.  The global minimum is at $x_d=105$ but $x_d=2000$ where there is 
no damping of oscillations for the Planck data gives a comparable fit, albeit with lower
oscillation amplitude $A_c$.   $\tilde A_s$ is an  effective normalization parameter defined in 
Eq.~(\ref{eqn:Astilde}) that determines the broadband observed CMB power in the
presence of $\tau$ and a step. }
 \label{table:representative_models}
\end{table}

For the step analysis,
the starting point is the canonical $c_s=1$ potential step where $C_1=C_3=0$.  As the
oscillatory features from $C_2$ dominate the fit to the Planck data, the other cases
are built from this model.  
In this case the step is described by $\{C_2, s_s, x_d \}$.
While we could directly minimize the $\chi^2$ in this joint cosmological and step parameter
space, the efficiency of the search is greatly improved by choosing parameters that 
are better aligned to the principle axes of the $\chi^2$ surface.  

The angular frequency of the oscillation changes with cosmological parameters at fixed 
$s_s$.  It is thus advantageous to replace $s_s$ with
\begin{align}
\theta_s &= \frac{s_s}{D_*},
\end{align}
where $D_*$ is the distance to recombination.  Note that the oscillations in $C_\ell$
are then described by sinusoids such as  $\sin(2 \ell \theta_s)$.
   Next, due to projection effects,
a fixed amplitude $C_2$ produces an oscillation in $C_\ell$ that decays as
$C_2 (\ell \theta_s)^{-1/2}$.   In Ref.~\cite{Adshead:2011jq} this scaling was approximately
accounted for in the curvature power spectrum description by introducing the amplitude parameter
\begin{align} \label{eqn:Ac_def}
A_c &= 3 C_2 \Bigg[\sqrt{\frac{s_s}{1\text{Gpc}}}\Bigg]^{-1} ;
\end{align}
we adopt this convention rather than the more orthogonal angular approach in order
to compare with the previous literature.
Note that the original version of the Planck collaboration analysis erroneously conflated this 
parameter with $C_2$ \cite{Ade:2013rta}.    

Finally, given
that the oscillations produce excess broadband power, we  use 
 the normalization parameter $\tilde A_s$ as defined in Eq.~(\ref{eqn:Astilde}).  For the
best fit models, this parameter rather than $A_s$ itself is nearly constant.
The optimized parameters are therefore
$\{\tilde{A}_s, n_s, \theta_A, \Omega_c h^2, \Omega_b h^2, \tau\}$
for the smooth cosmology and $\{ A_c, \theta_s, x_d \}$ for the step. 
The minimum $\chi^2$ potential step model with $c_s=1$ is given in 
Tab.~\ref{table:representative_models} (GSR1 column) and represents an improvement of 
$\Delta\chi^2 = -14.0$ over the smooth model of Tab.~\ref{table:smooth_parameters}.
For reference we also show here the best fit model at $x_d=2000$, where the oscillations
are undamped all the way to the maximum of $\ell=2500$ for Planck.   Note that most
of the improvement due to the step remains.  We also repeat the minimization for the
GSR0 approximation used in previous treatments for comparison.  Note that after adjusting
cosmological parameters, steps in either approximation fit equally well but the recovery of
such parameters would be biased by using the less accurate GSR0 approximation.

For the arbitrary sound speed warp and potential step models, $C_1$ and $C_3$ are  set consistently with the step amplitude $\{ b_T, b_V \}$ and slow roll parameters
after the step $\{c_{sa},\epsilon_{Ha}\}$ through Eq.~(\ref{eqn:CW}).     These parameters mainly change the power spectrum around $\ell \sim 1/\theta_s$ and hence produce only small changes in the $\chi^2$ due to the limitations of cosmic 
variance.  

	We therefore keep the other parameters fixed to the values of $x_d=105$ model listed in Tab.~\ref{table:representative_models} when considering the additional freedom in these models.   Given a fixed $C_2$, which fixes the amplitude of the step, this freedom is parameterized by
$c_{sa}$, the sound speed after the step. For warp step, the best fit is given by
\begin{align}
	c_{sa} &= 0.70, \nonumber\\
	C_1 &= -0.70, \nonumber\\
	C_3 &= -0.37, \nonumber\\
	\Delta \chi^2 &= -15.2, \quad (\text{warp}),
\end{align}
and this corresponds to a $\Delta \chi^2 = -1.2$ improvement over the potential step model at $c_{sa}=1$. For low sound speed potential step models, the $c_{sa} \to 0$ limit provides the best fit
\begin{align}
	c_{sa} &\rightarrow 0 ,  \nonumber\\
	C_1 &=0, \nonumber\\
	C_3 &= -0.03, \nonumber\\
	\Delta \chi^2 &= -14.1, \quad (\text{potential}).
\end{align}
Given the additional parameter $c_{sa}$, neither improvement is statistically significant.

\vfill 
\break
\bibliography{DBIBI2}
\vfill

\end{document}